\documentclass[pra,twocolumn,amsmath,longbibliography]{revtex4-1}

\usepackage{graphicx}
\renewcommand{\d}{{d}}

\begin{document}

\title{Interaction potential for NaCs for ultracold scattering and spectroscopy}

\author{Samuel G. H. Brookes}

\affiliation{Joint Quantum Centre (JQC) Durham-Newcastle, Department of
Chemistry, Durham University, South Road, Durham DH1 3LE, United Kingdom}

\author{Jeremy M. Hutson}
\affiliation{Joint Quantum Centre (JQC) Durham-Newcastle, Department of
Chemistry, Durham University, South Road, Durham DH1 3LE, United Kingdom}
\email{J.M.Hutson@durham.ac.uk}

\date{\today}


\begin{abstract}
We obtain the interaction potential for NaCs by fitting to experiments on ultracold scattering and spectroscopy in optical tweezers. The central region of the potential has been accurately determined
from Fourier-Transform spectroscopy at higher temperatures, 
so we focus on adjusting the long-range and short-range parts. We use coupled-channel calculations of binding energies and wave functions to understand the nature of the molecular states observed in ultracold spectroscopy, and of the state that causes the Feshbach resonance used to create ultracold NaCs molecules. We elucidate the relationships between the experimental quantities and features of the interaction potential. We establish the combinations of experimental quantities that determine particular features of the potential. We find that the long-range dispersion coefficient $C_6$ must be increased by about 0.9\% to 3256(1) $E_\textrm{h} a_0^6$ to fit the experimental results. We use coupled-channel calculations on the final potential to predict bound-state energies and resonance positions.
\end{abstract}

\maketitle

\section{Introduction}

Ultracold polar molecules have many potential applications, ranging from precision measurement
\cite{Zelevinsky:mass-ratio:2008, Hudson:2011, Salumbides:2011, Salumbides:2013, Schiller:2014, ACME:2014,
Hanneke:2016, Cairncross:2017, Borkowski:2018, ACME:2018, Borkowski:2019}, quantum
simulation~\cite{Barnett:2006, Micheli:2006, Buechler:2007, Macia:2012, Manmana:2013, Gorshkov:2013} and
quantum information processing~\cite{DeMille:2002, Yelin:2006, Zhu:2013, Herrera:2014, Ni:2018,
Sawant:qudit:2020, Hughes:2020} to state-resolved chemistry~\cite{Krems:PCCP:2008, Bell:2009,
Ospelkaus:react:2010, Dulieu:2011a, Balakrishnan:2016, Hu:2019}. A very important class of ultracold
molecules are the alkali-metal diatomic molecules; these are usually produced by association of
pairs of ultracold atoms, by magnetoassociation or photoassociation, followed by coherent optical
transfer to the ground rovibronic state. The ground-state molecules produced in this way include
KRb~\cite{Ni:KRb:2008, Voges:NaK:2020}, Cs$_2$~\cite{Danzl:v73:2008, Danzl:ground:2010},
Rb$_2$~\cite{Lang:cruising:2008},
RbCs~\cite{Takekoshi:RbCs:2014, Molony:RbCs:2014}, NaK~\cite{Park:NaK:2015, Seesselberg:2018,
Yang:K_NaK:2019}, NaRb~\cite{Guo:NaRb:2016}, NaLi~\cite{Rvachov:2017} and
NaCs~\cite{Cairncross:2021}.

A particular success in the last few years has been the production of ultracold NaCs molecules in optical tweezers. Configurable arrays of polar molecules in tweezers offer many possibilities for studying few-body physics involving dipolar species and constructing designer Hamiltonians for quantum logic and quantum simulation. In 2018, Liu et al.\ \cite{Liu:NaCs:2018} succeeded in loading one atom each of Na and Cs into a single optical tweezer, and photoassociated them to form a single electronically excited NaCs molecule in the tweezer.  Liu et al.\ \cite{Liu:NaCs:2019} measured the binding energy of the least-bound triplet state of NaCs by two-photon Raman spectroscopy. Hood et al.\ \cite{Hood:NaCs:2020} measured interaction shifts for flipping the spin of one or both atoms in the tweezer, and located magnetically tunable Feshbach resonances in an excited spin channel. They used these measurements to model the interaction using multichannel quantum defect theory (MQDT). Zhang et al.\ \cite{Zhang:NaCs:2020} located an s-wave Feshbach resonance in the lowest spin channel, allowing them to form a single NaCs molecule in the tweezer by magnetoassociation. Yu et al.\ \cite{Yu:NaCs:2021} used a different route to form a single NaCs molecule in the tweezer by coherent Raman transfer. Most recently, Cairncross et al.\ \cite{Cairncross:2021} transferred a molecule formed by magnetoassociation to the absolute ground state by a coherent Raman process.

Studies of ultracold molecule formation typically need close collaboration between experiment and theory. Initial experiments identify properties of the system that can be used to determine an initial interaction potential. The interaction potential is then used to predict new experimental properties. Once these are measured, they are used to refine the interaction potential, and the process repeats. The studies of NaCs in tweezers have followed this cycle several times. In the process, we have learnt a considerable amount, both about the specific system and more generally about the ways in which experimental properties are influenced by features of the interaction potential. The purpose of the present paper is to present the fitted potential for Na+Cs, describe its relationships to experimental observables, and explain the insights that have been gained. Accurate interaction potentials have applications not only for ultracold molecules but also for precise control of atomic collisions, for example in studies of Efimov physics \cite{Huang:2nd-Efimov:2014} and quantum droplet formation \cite{Guo:NaRb:2022}.

The structure of this paper is as follows. Section \ref{sec:theory} describes the underlying theory and the methods used in the present work.
Section \ref{sec:obs} describes the measured quantities from ultracold scattering and spectroscopy, the wave functions of the underlying weakly bound states, and their relationship to the singlet and triplet potential curves. Section \ref{sec:fitting} describes our procedure for fitting potential parameters, with a focus on how each parameter is related to and constrained by the measured quantities. Section \ref{sec:predict} describes the near-threshold bound states calculated for our final interaction potential, and the resulting scattering properties, including predictions for additional resonances. It compares additional measurements for p-wave and d-wave resonances and gives assignments for the states involved. Finally, Section \ref{sec:conc} summarizes our conclusions and the insights gained from the present work.

\section{Theoretical Methods} \label{sec:theory}

\subsection{Atomic states}

The Hamiltonian for an alkali-metal atom $X$ in its ground $^2$S state may be written
\begin{equation}
\hat h_X = \zeta_X \hat \imath_X \cdot \hat s_X + g_{S,X} \mu_{\rm B}
B \, \hat s_{X,z} + g_{n,X} \mu_{\rm B} B \, \hat \imath_{z,X},
\label{eq:h-hat}
\end{equation}
where $\zeta_X$ is the hyperfine coupling constant, $\hat s_X$ and $\hat \imath_X$ are the operators for the electron and nuclear spin, respectively, and $\hat s_{z,X}$ and $\hat \imath_{z,X}$ represent their $z$-components along an axis defined by the external magnetic field $B$ \footnote{We follow the convention of using lower-case letters for operators and quantum numbers of individual atoms, and upper-case letters for those of the diatomic molecule or colliding pair of atoms.}. The constants $g_{S,X}$ and $g_{n,X}$ are the electron and nuclear $g$-factors and $\mu_{\rm B}$ is the Bohr magneton. The numerical values are taken from Steck's compilations \cite{Steck:Na, Steck:Cs}.

\begin{figure}[tbp]
 \centering
    \includegraphics[width=\columnwidth]{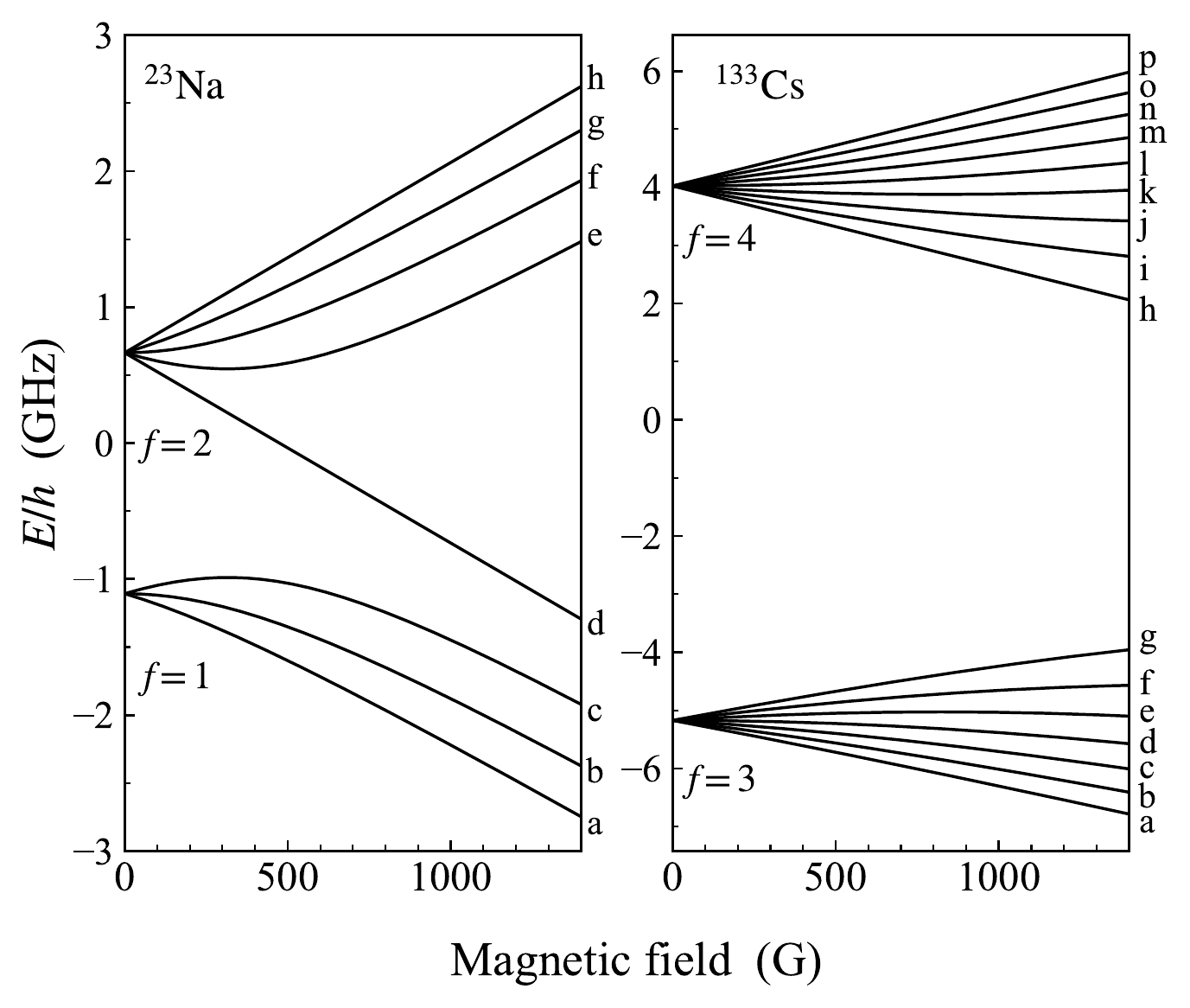}
    \caption{Breit-Rabi plots showing the hyperfine structure and Zeeman splitting for  $^{23}$Na and $^{133}$Cs atoms. The zero of energy is the hyperfine centroid in each case. Each state is identified by a Roman letter in alphabetic order from the lowest, which is designated a.}
    \label{fig:breit}
\end{figure}

The nuclear spin is $i=3/2$ for $^{23}$Na and $i=7/2$ for $^{133}$Cs. These are the only stable isotopes for each element, so in the following we drop the mass numbers. The hyperfine splitting at zero field is $(i+\frac{1}{2})\zeta_X$ and is approximately 1.77~Gz for Na and 9.19~GHz for Cs. Because of these differences, the free atoms have quite different different Zeeman structures, as shown in Fig.\ \ref{fig:breit}.

At low fields the atomic states may be labeled with $f=i\pm\frac{1}{2}$ and its projection $m_f$ onto the axis of the magnetic field. However, at higher fields the magnetic field mixes states of different $f$, particularly for Na. Here we label the states alphabetically in increasing order of energy, with Roman letters from a to h for Na and from a to p for Cs, as shown in Fig.\ \ref{fig:breit}. In each case the highest-energy state is spin-stretched, with $f=m_f=i+\frac{1}{2}$.

We label a state of an atom pair with two letters, with Na first: for example, ha indicates that Na is in its uppermost state and Cs in its lowest. The \emph{threshold} for a particular pair state is the energy of the separated atom pair at the appropriate magnetic field.
There are $128 = (3 + 5) \times (7 + 9)$ of these thresholds, but no more than 16 for a particular value of $M_F = m_{f,\textrm{Na}} + m_{f,\textrm{Cs}}$, which is a nearly conserved quantity in a magnetic field.

\subsection{Two-atom Hamiltonian}

When two alkali-metal atoms in their ground $^2$S states approach one another, their electron spins $s_1=s_2=\frac{1}{2}$ couple to form either a singlet state $X{}^1\Sigma^+$ with total electron spin $S=0$ or a triplet state $a{}^3\Sigma^+$ with $S=1$. Their interaction is governed mostly by the electrostatic potential curves $V_0(R)$ and $V_1(R)$ for the singlet and triplet states, respectively, but there are also small spin-dependent terms as described below.

The Hamiltonian for an interacting pair of atoms may be written
\begin{equation}
\frac{\hbar^2}{2\mu} \left(-R^{-1} \frac{d^2}{dR^2} R + \frac{\hat{L}^2}{R^2} \right) + \hat h_1 + \hat h_2 + \hat V(R),
\label{eq:SE}
\end{equation}
where $R$ is the internuclear distance, $\mu$ is the reduced mass and $\hat L$ is the operator for the end-over-end angular momentum of the two atoms about one another.

\begin{figure}[tb]
 \centering
    \includegraphics[width=1.0\columnwidth]{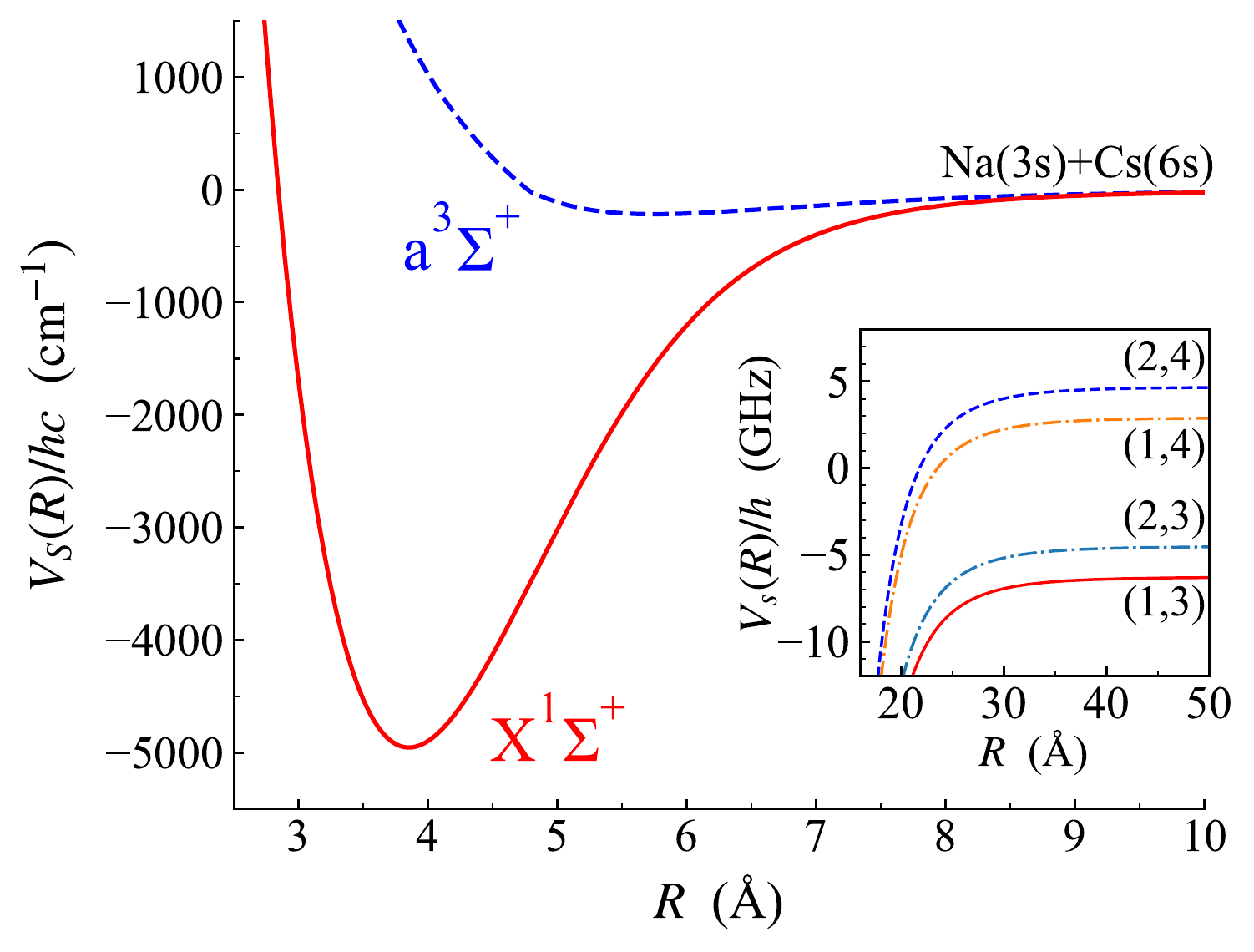}
    \caption{Potential curves of Docenko et al.\ \cite{Docenko:2006} for the $X{}^1\Sigma^+$ and $a{}^3\Sigma^+$ states of NaCs. The inset shows an expanded view of the zero-field hyperfine structure at long range, with thresholds labeled ($f_{\textrm{Na}},f_{\textrm{Cs}}$) and energies shown relative to the hyperfine centroid.
    }
    \label{fig:potentials}
\end{figure}

The interaction between the atoms is described by the interaction operator, which for a pair of alkali-metal atoms takes the form
\begin{equation}
{\hat V}(R) = \hat V^{\rm c}(R) + \hat V^\textrm{d}(R).
\label{eq:V-hat}
\end{equation}
Here $\hat V^{\rm c}(R)=V_0(R)\hat{\cal{P}}^{(0)} + V_1(R)\hat{\cal{P}}^{(1)}$ is an isotropic potential operator that accounts for the potential energy curves $V_0(R)$ and $V_1(R)$ for the singlet and triplet states. The singlet and triplet projectors $\hat{\cal{P}}^{(0)}$ and $\hat{ \cal{P}}^{(1)}$ project onto subspaces with $S=0$ and 1 respectively. Figure~\ref{fig:potentials} shows the two potential energy curves for NaCs. The functional forms used for these are described in Section \ref{sec:curves}.

The term $\hat V^\textrm{d}(R)$ describes the dipole-dipole interaction between the magnetic moments of the electrons at long range, together with terms due to 2nd-order spin-orbit coupling at short range. This makes only small contributions for the experimental observables that we fit to in the present paper, but it is important for some of the predicted observables described in Section \ref{sec:predict}. It is described in Appendix \ref{app:SO}.

\subsection{Calculations of bound states and scattering}

We carry out calculations of both bound states and scattering using coupled-channel methods, as described in Appendix \ref{app:cc}. The total wave function is expanded in a complete basis set of functions for electron and nuclear spins and end-over-end rotation, producing a set of coupled differential equations that are solved by propagation with respect to the internuclear distance $R$. The coupled equations are identical for bound states and scattering, but the boundary conditions are different.

Scattering calculations are performed with the \textsc{molscat} package \cite{molscat:2019, mbf-github:2020}. Such calculations produce the scattering matrix $\boldsymbol{S}$, for a single value of the collision energy and magnetic field each time. The complex s-wave scattering length $a(k_0)$ is obtained from the diagonal element of $\boldsymbol{S}$ in the incoming channel, $S_{00}$, using the identity \cite{Hutson:res:2007}
\begin{equation}
a(k_0) = \frac{1}{ik_0} \left(\frac{1-S_{00}(k_0)}{1+S_{00}(k_0)}\right),
\end{equation}
where $k_0$ is the incoming wavenumber, related to the collision energy $E_\textrm{coll}$ by $E_\textrm{coll}=\hbar^2k_0^2/(2\mu)$. The scattering length $a(k_0)$ becomes constant at sufficiently low $E_\textrm{coll}$, with limiting value $a$. In the present work, s-wave scattering lengths are calculated at $E_\textrm{coll}/k_\textrm{B} = 1$~nK, which is low enough to neglect the dependence on $k_0$.

A zero-energy Feshbach resonance occurs where a bound state of the atomic pair (diatomic molecule) crosses a scattering threshold as a function of applied field. At the lowest threshold, or in the absence of inelastic processes, the scattering length is real. Near a resonance, $a(B)$ passes through a pole, and is approximately
\begin{equation}
a(B) = a_\textrm{bg} \left( 1 - \frac{\Delta}{B-B_\textrm{res}}\right),
\label{eq:res}
\end{equation}
where $B_\textrm{res}$ is the position of the resonance, $\Delta$ is its width, and $a_\textrm{bg}$ is a slowly varying background scattering length. In the presence of inelastic processes, $a(B)$ is complex and the pole is replaced by an oscillation \cite{Hutson:res:2007}. \textsc{molscat} can converge on Feshbach resonances automatically and characterize them to obtain $B_\textrm{res}$, $\Delta$ and $a_\textrm{bg}$ (and the additional parameters needed in the presence of inelasticity) as described in ref.\ \citenum{Frye:resonance:2017}.

Coupled-channel bound-state calculations are performed using the packages \textsc{bound} and \textsc{field} \cite{bound+field:2019, mbf-github:2020}, which converge upon bound-state energies at fixed field, or bound-state fields at fixed energy, respectively. The methods used are described in ref.\ \citenum{Hutson:CPC:1994}. Once bound states have been located, their wave functions may be obtained by back-substitution using matrices saved from the original propagation \cite{THORNLEY:1994}. Alternatively, the expectation value of any operator may be calculated by finite differences, without requiring explicit wave functions \cite{Hutson:expect:88}. This capability is used here to calculate overall triplet fractions for bound states.

Zero-energy Feshbach resonances can be fully characterized using \textsc{molscat} as described above. However, if only the position of the resonance is needed, it is more convenient simply to run \textsc{field} at the threshold energy to locate the magnetic field where the bound state crosses threshold.

A key capability of both \textsc{molscat} and \textsc{field}, used in the present work, is automated convergence of any one parameter in the interaction potential to reproduce a single observable quantity, such as a bound-state energy, scattering length, or resonance position. This uses the same algorithms as are used to converge on such quantities as a function of external field \cite{Hutson:CPC:1994, Frye:resonance:2017}.

In the present work, the coupled equations for both scattering and bound-state calculations are solved using the fixed-step symplectic log-derivative propagator of Manolopoulos and Gray \cite{MG:symplectic:1995} from $R_\textrm{min}=4\ a_0$ to $R_\textrm{mid}=30\ a_0$, with an interval size of $0.002\ a_0$, and the variable-step Airy propagator of Alexander and Manolopoulos \cite{Alexander:1987} between $R_\textrm{mid}$ and $R_\textrm{max}=10,000\ a_0$. The exception to this is calculations used to plot wave functions, which use the fixed-step log-derivative propagator of Manolopoulos \cite{Manolopoulos:1986, THORNLEY:1994}.

\subsection{Basis sets for angular momentum} \label{sec:basis}

To carry out coupled-channel calculations, we need a basis set that spans the space of electron and nuclear spins and of relative rotation. We do not require a basis set where the atomic Hamiltonians $\hat{h}_1$ and $\hat{h}_2$ are diagonal, because \textsc{molscat} transforms the solutions of the coupled equations into an asymptotically diagonal basis set before applying scattering boundary conditions.

There are 5 sources of angular momentum for an interacting pair of alkali-metal atoms: the electron spins $s_1$ and $s_2$, the nuclear spins $i_1$ and $i_2$, and the rotational angular momentum $L$. These may be coupled together in several different ways, and different coupling schemes are useful when discussing different aspects of the problem. The separated atoms are conveniently represented by quantum numbers $(s,i)f,m_f$, where the notation $(a,b)c$ indicates that $c$ is the resultant of $a$ and $b$ and $m_c$ is the projection of $c$ onto the $z$ axis. Conversely, the molecule at short range (and low field) is better represented by $S$ and the total nuclear spin $I$, together with their resultant $F$ and its projection $M_F$.
In the present work, we carry out coupled-channel calculations in two different basis sets. The first is
\begin{equation}
|(s_\textrm{Na},i_\textrm{Na})f_\textrm{Na},m_{f,\textrm{Na}}\rangle
|(s_\textrm{Cs},i_\textrm{Cs})f_\textrm{Cs},m_{f,\textrm{Cs}}\rangle
|L,M_L\rangle,
\label{eq:basis-part}
\end{equation}
which we term the \emph{coupled-atom} basis set. The second is
\begin{equation}
|\left((s_\textrm{Na},s_\textrm{Cs})S,
(i_\textrm{Na},i_\textrm{Cs})I\right)FM_F\rangle
|L,M_L\rangle,
\label{eq:basis-sif}
\end{equation}
which we term the $SIF$ basis set.
The only conserved quantities in a magnetic field are $M_\textrm{tot} = m_{f,\textrm{Na}} + m_{f,\textrm{Cs}} + M_L = M_F + M_L$ and parity $(-1)^L$. We take advantage of this to perform calculations for each $M_\textrm{tot}$ and parity separately. In each calculation, we include all basis functions of the required $M_\textrm{tot}$ and parity for $s_\textrm{Na}=s_\textrm{Cs}=\frac{1}{2}$, $i_\textrm{Na}=\frac{3}{2}$ and $i_\textrm{Cs}=\frac{7}{2}$, subject to the limitation $L\le L_\textrm{max}$. In most of the calculations in the present work, $L_\textrm{max}=0$, except that we use $L_\textrm{max}=1$ for calculations of p-wave states and resonances in Section \ref{sec:res-p} and $L_\textrm{max}=2$ for the calculations in Section \ref{sec:res-s}.


\subsection{Singlet and triplet potential curves} \label{sec:curves}

Our starting points for fitting the interaction potentials are the singlet and triplet potential curves of Docenko et al.\ \cite{Docenko:2006}, shown in Fig.\ \ref{fig:potentials}. These were fitted to extensive Fourier-transform (FT) spectra involving vibrational levels up to $v=83$ in the singlet state, which has a total of 88 levels, and up to $v=21$ in the triplet, which has 25. These curves give an excellent representation of the levels they were fitted to, but their behavior at higher energies depends sensitively on how they are extrapolated, and they do not reproduce the near-threshold states important for ultracold scattering.

In a central region from $R_{\textrm{SR},S}$ to $R_{\textrm{LR},S}$, with $S=0$ for the singlet and $S=1$ for the triplet, each curve is represented as a finite power series in a nonlinear function $\xi_S$ that depends on the internuclear separation $R$,
\begin{equation}
V_{\textrm{mid},S}(R) = \sum_{i=0}^{n_S} a_{i,S} \xi_S^i(R),
\label{eq:mid}
\end{equation} where
\begin{equation}
\xi_S(R)= \frac{R-R_{\textrm{m},S}}{R+b_S R_{\textrm{m},S}}.
\label{eq:xi}
\end{equation}
The quantities $a_{i,S}$ and $b_S$ are fitting parameters, and $R_{\textrm{m},S}$ is chosen to be near the equilibrium distance for the state concerned. The values of the parameters fitted to FT spectroscopy for NaCs are given in Tables 1 and 2 of ref.\ \citenum{Docenko:2006}; the values $R_\textrm{SR,0} = 2.8435$~\AA\ and $R_\textrm{SR,1} = 4.780$~\AA, which specify the minimum distance at which the power-series expansion is used for each state, are particularly important for the present work.

At long range ($R > R_{\textrm{LR},S}$), the potentials are
\begin{equation}
\begin{split}
V_{\textrm{LR},S}(R) = -C_6/R^6 - C_8/R^8 - C_{10}/R^{10} \\
-(-1)^S V_\textrm{ex}(R),
\end{split}
\label{eq:lr}
\end{equation}
where the dispersion coefficients $C_n$ are common to both potentials. The long-range matching points are chosen as $R_{\textrm{LR,0}} = R_{\textrm{LR,1}} = 10.2$~\AA.
The exchange contribution is \cite{Smirnov:1965}
\begin{equation}
V_\textrm{ex}(R) = A_{\rm ex} R^\gamma \exp(-\beta R),
\end{equation}
where $a_0$ is the Bohr radius. It makes an attractive contribution for the singlet and a repulsive contribution for the triplet. The value of $C_6$ used by Docenko et al.\ \cite{Docenko:2006} was fixed at the theoretical value of Derevianko et al.\ \cite{Derevianko:2001}, while $C_8$, $C_{10}$ and $A_{\rm ex}$ were fitting parameters. The mid-range potentials are adjusted to match the long-range potentials at $R_{\textrm{LR},S}$ by setting the constant terms $a_{0,S}$ in Eq.\ \ref{eq:mid} as required.

Lastly, the potentials are extended to short range ($R < R_{\textrm{SR},S}$) with simple repulsive terms,
\begin{equation}
V_{\textrm{SR},S}(R) = A_{\textrm{SR},S} + B_{\textrm{SR},S} /R^{N_S},
\label{eq:sr}
\end{equation}
where $A_{\textrm{SR},S}$ is chosen so that $V_{\textrm{SR},S}$ and $V_{\textrm{mid},S}$ match at $R_{\textrm{SR},S}$. In the present work, $B_{\textrm{SR},S}$ is chosen to match the derivative of these two functions. However, this latter constraint was not applied in ref.\ \citenum{Docenko:2006}, producing discontinuities in the derivatives of the potential curves at $R_{\textrm{SR},S}$.

\section{Results and Discussion} \label{sec:results}

\subsection{Observables from ultracold scattering and spectroscopy} \label{sec:obs}

The recent experimental studies on Na+Cs in tweezers \cite{Liu:NaCs:2019, Zhang:NaCs:2020, Hood:NaCs:2020, Yu:NaCs:2021, Cairncross:2021} have measured a number of quantities that could be used in fitting potential curves. Each observable is associated with one or more molecular bound states of a particular spin character. In this section we consider each observable quantity and the nature of the corresponding state, in order to understand how the observable depends on features of the singlet and triplet potential curves. The calculations in this section are based on `lightly-fitted' potential curves, with approximately correct scattering lengths. Calculations based on the final potential would be visually almost identical.

\subsubsection{General features of near-threshold states}

The near-threshold states that are important in studies of ultracold molecules and ultracold collisions  are typically bound by less than a few GHz. Their wave fuctions extend several nm to distances where hyperfine coupling is stronger that the spacing between the singlet and triplet curves. This long-range region is shown as an inset in Figure \ref{fig:potentials}. Each curve represents a different zero-field hyperfine threshold, labeled ($f_{\textrm{Na}},f_{\textrm{Cs}}$). For an interaction potential of the form $-C_6/R^6$ at long range, the bound states below each threshold are located within `bins' given by multiples of an energy scale $\bar{E}=\hbar^2/(2\mu \bar{a}^2)$ \cite{Gao:2000}, where $\bar{a}$ is the mean scattering length \cite{Gribakin:1993} and depends only on $C_6$ and $\mu$. For NaCs, $\bar{a} = 59.17\ a_0$ and $\bar{E} = 26.30$~MHz. The first (top) bin is 36.1$\bar{E} = 950$~MHz deep, implying that the top (least-bound) bound state lies 0 to 950~MHz below its threshold; the position of the state within the bin is governed by the actual scattering length $a$, which differs for different thresholds. The least-bound state is designated $n=-1$. The second and third bins extend to depths of $249\bar{E}$ and $796\bar{E}$, so the second and third bound states (with $n=-2$ and $-3$) lie between 950 MHz and 6.6 GHz and between 6.6 and 21~GHz below threshold, respectively. We focus here on states with binding energies within the three uppermost bins; accurately modeling of this region of the potential is crucial for obtaining reliable scattering lengths and resonance positions, among other properties.

\subsubsection{Binding energy of the absolute ground state}

Cairncoss et al.\ \cite{Cairncross:2021} have measured the energy of the absolute ground state of NaCs, initially with respect to the near-threshold state formed by magnetoassociation. After correcting for hyperfine and Zeeman effects and the binding energy of the near-threshold state, they infer that the binding energy $E_{00}$ of the lowest rovibrational level of the singlet state, relative to the hyperfine centroid of free atoms, is 147,044.63(11) GHz.

This state is located thousands of cm$^{-1}$ below the minimum of the triplet state, so singlet-triplet mixing is negligible. Its binding energy is sensitive only to the singlet curve. Its wave function is tightly confined around the minimum of the singlet curve near 3.85~\AA, and the zero-point energy is very well determined by the FT spectra, so it is mostly sensitive to the well depth of the singlet curve.

\subsubsection{Binding energy of least-bound pure triplet state}

The binding energy of the least-bound state in the hp channel, $E_{-1}^\textrm{hp}$, has been measured by Liu et al.\ \cite{Liu:NaCs:2019} and refined by Hood et al.\ \cite{Hood:NaCs:2020}. This channel corresponds to $(f,m_f)=(2,2)$ for Na and (4,4) for Cs. Both these states are spin-stretched, with $f=m_f=s+i$, so states that lie in the hp channel are pure triplet in character. The binding energy of the state, relative to the hp threshold, is 297.6(1) MHz at 8.8 G.

The binding energy $E_{-1}^\textrm{hp}$ is sensitive only to the triplet curve. It is also very closely related to the triplet scattering length $a_\textrm{t}$, with only slight sensitivity to the dispersion coefficient $C_6$ and even less to $C_8$ and $C_{10}$.

\subsubsection{Binding energy of least-bound state in ha channel}

\begin{figure}[tbp]
 \centering
    \includegraphics[width=1\columnwidth]{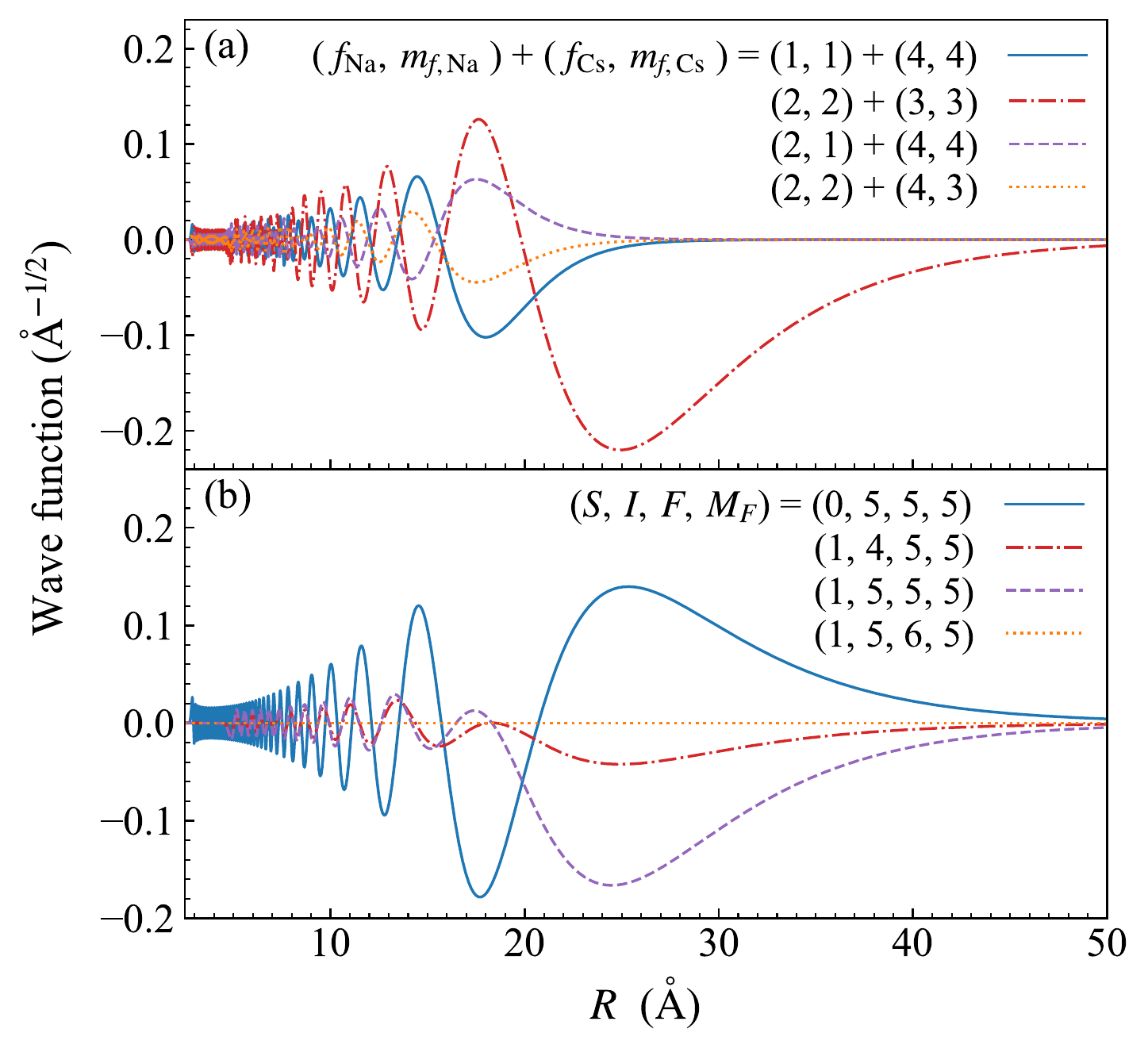}
    \caption{Components of the wave function for the least-bound state in the ha channel, shown in both the (a) coupled-atom and (b) $SIF$ representations. Components in all four contributing channels are plotted in each case.}
    \label{fig:ha_wf}
\end{figure}

\noindent Yu et al.\ \cite{Yu:NaCs:2021} have measured the binding energy of the least-bound state in the ha channel, $E_{-1}^\textrm{ha}$, with respect to the ha threshold. The binding energy is 770.1969(2) MHz at $B=8.83$~G.

The ha channel corresponds to $(f,m_f)=(2,2)$ for Na and (3,3) for Cs, so $M_F=5$. Since there are 4 atomic pair states with $M_F=5$, which are mixed by the interaction potential, this state has a mixture of singlet and triplet character. To quantify this, Fig.\ \ref{fig:ha_wf} shows the components of the wave function for this state. In the coupled-atom representation, the main contribution is provided by the ha channel, with smaller contributions arising from the other three channels with $M_F=5$. In the $SIF$ representation, there are similar contributions from singlet and triplet channels. The overall triplet fraction obtained from the expectation value of the triplet projector $\hat{\cal{P}}_1$ is 49.7\%.

The binding energy $E_{-1}^\textrm{ha}$ is approximately equally sensitive to the singlet and triplet curves. It is closely related to the scattering length in the ha channel. However, since the triplet scattering length is determined independently by $E_{-1}^\textrm{hp}$, the role of $E_{-1}^\textrm{ha}$ is to provide information on the singlet scattering length $a_\textrm{s}$.

\begin{figure}[tbp]
 \centering
    \includegraphics[width=1\columnwidth]{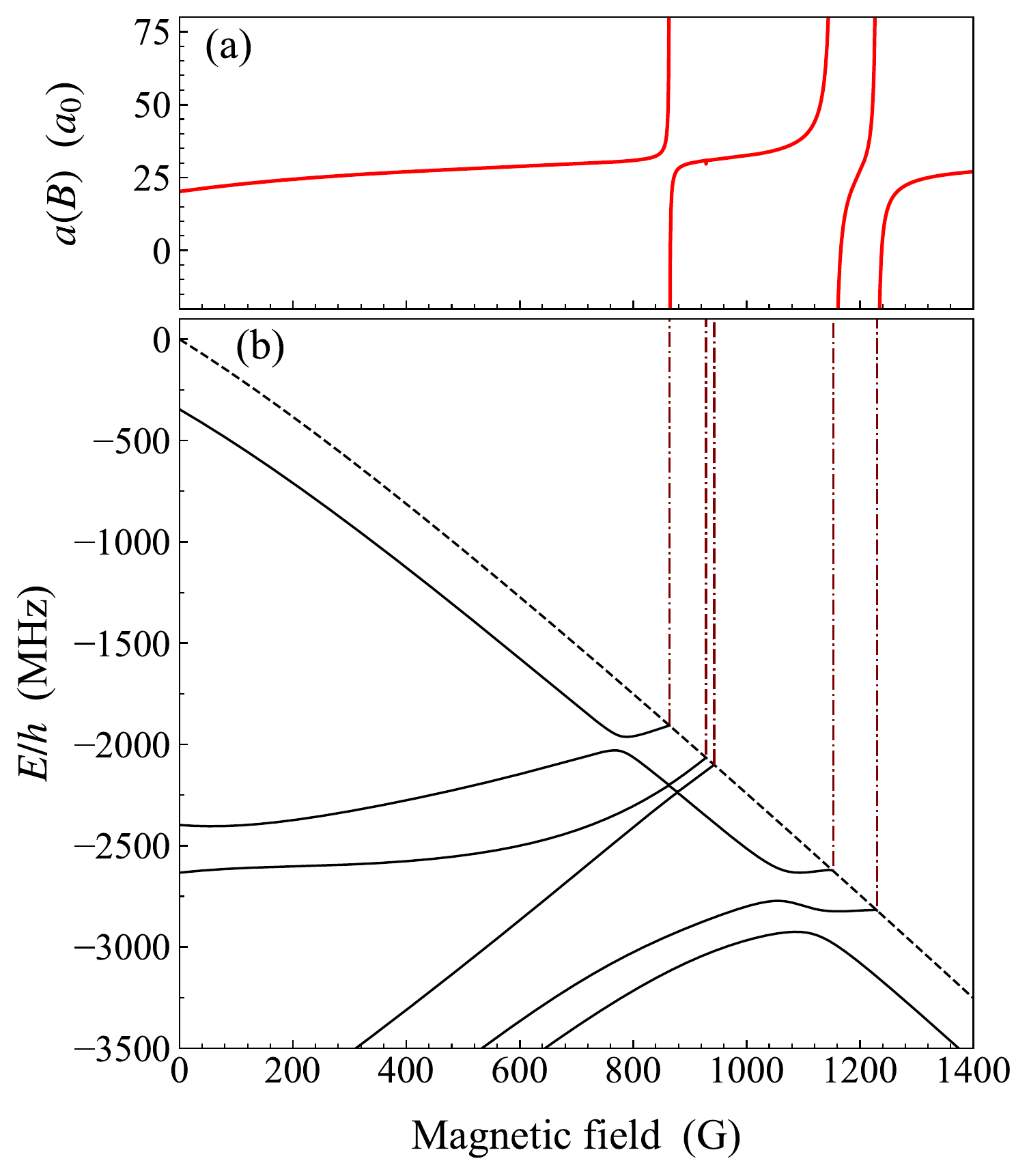}
    \caption{(a) The calculated s-wave scattering length in the aa channel as a function of magnetic field. (b) Energies of weakly bound s-wave molecular states with $M_F=4$ (solid lines) and the aa threshold (dashed line). The zero of energy is the zero-field threshold energy. Feshbach resonances occur where bound states cross threshold and are indicated by vertical lines extending up to the corresponding position on the plot of the scattering length.
    }
    \label{fig:aa_states}
\end{figure}

\begin{figure}[tbp]
 \centering
    \includegraphics[width=1\columnwidth]{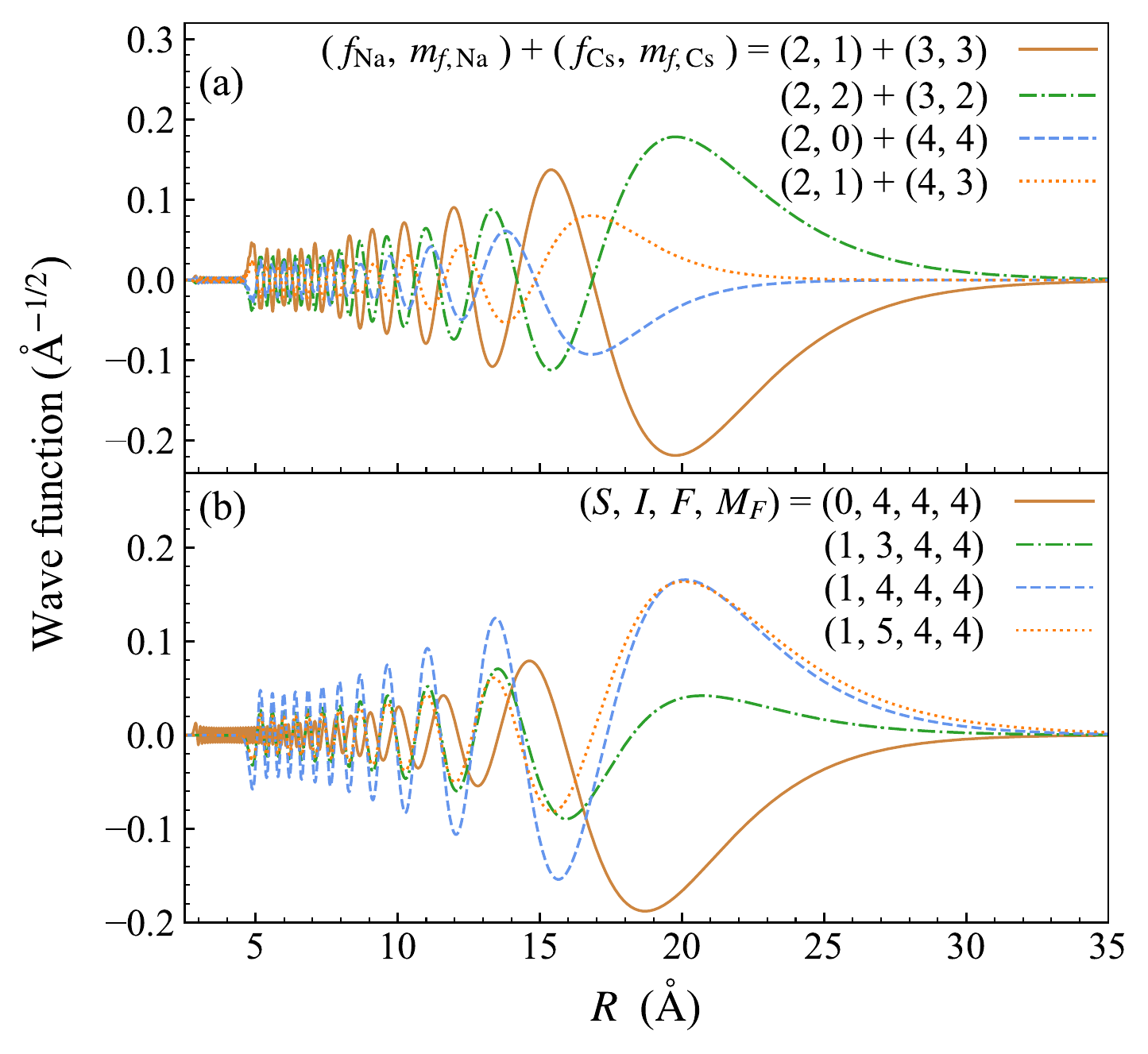}
    \caption{Components of the wave function at zero field for the state responsible for the resonance near 864~G in the aa channel, shown in both the (a) coupled-atom and (b) $SIF$ representations. Components in the four most prominent channels are plotted in each case.}
    \label{fig:aa_wf}
\end{figure}

\subsubsection{Position of Feshbach resonance in aa channel}

Zhang et al.\ \cite{Zhang:NaCs:2020} have observed a strong s-wave resonance in the lowest hyperfine channel at 864.11(5)~G and used it to form NaCs molecules by magnetoassociation. The atoms collide at the aa threshold, corresponding to $(f,m_f)=(1,1)+(3,3)$ at low field. The resonance position is designated $B_\textrm{res}^\textrm{aa}$.

Figure \ref{fig:aa_states} shows the pattern of s-wave bound states below the aa threshold as a function of magnetic field, obtained from coupled-channel bound-state calculations, together with the calculated scattering length. The bound state originating at $-400$ MHz and running parallel to the aa threshold has the same spin character (i.e. the same spin quantum numbers) as the aa threshold. The resonance near 864~G occurs when this state is pushed up and across the threshold by a more deeply bound state through an avoided crossing.

The more deeply bound state originates from $-2450$ MHz below the aa threshold at zero field. Its depth and behavior with magnetic field ultimately determine the location and nature of the resulting resonance. The components of its wave function at zero field are plotted in Fig.\ \ref{fig:aa_wf}. In the coupled-atom representation, the dominant components are from channels corresponding to $(f_\textrm{Na},f_\textrm{Cs})=(2,3)$ (solid brown and dot-dashed green curves). The calculated zero-field binding energy is 4220 MHz below the (2,3), threshold, indicating that the state corresponds to $n=-2$. Because of this, the wave function is concentrated at significantly shorter range than those for the least-bound states in Fig.\ \ref{fig:ha_wf}. The components of the wave function in the $SIF$ representation are shown in Fig.\ \ref{fig:aa_wf}(b). There are significant contributions from both singlet and triplet channels. The overall triplet fraction is 69.5\%.

\subsubsection{Position of Feshbach resonance in cg channel}

\begin{figure}[tbp]
 \centering
    \includegraphics[width=1\columnwidth]{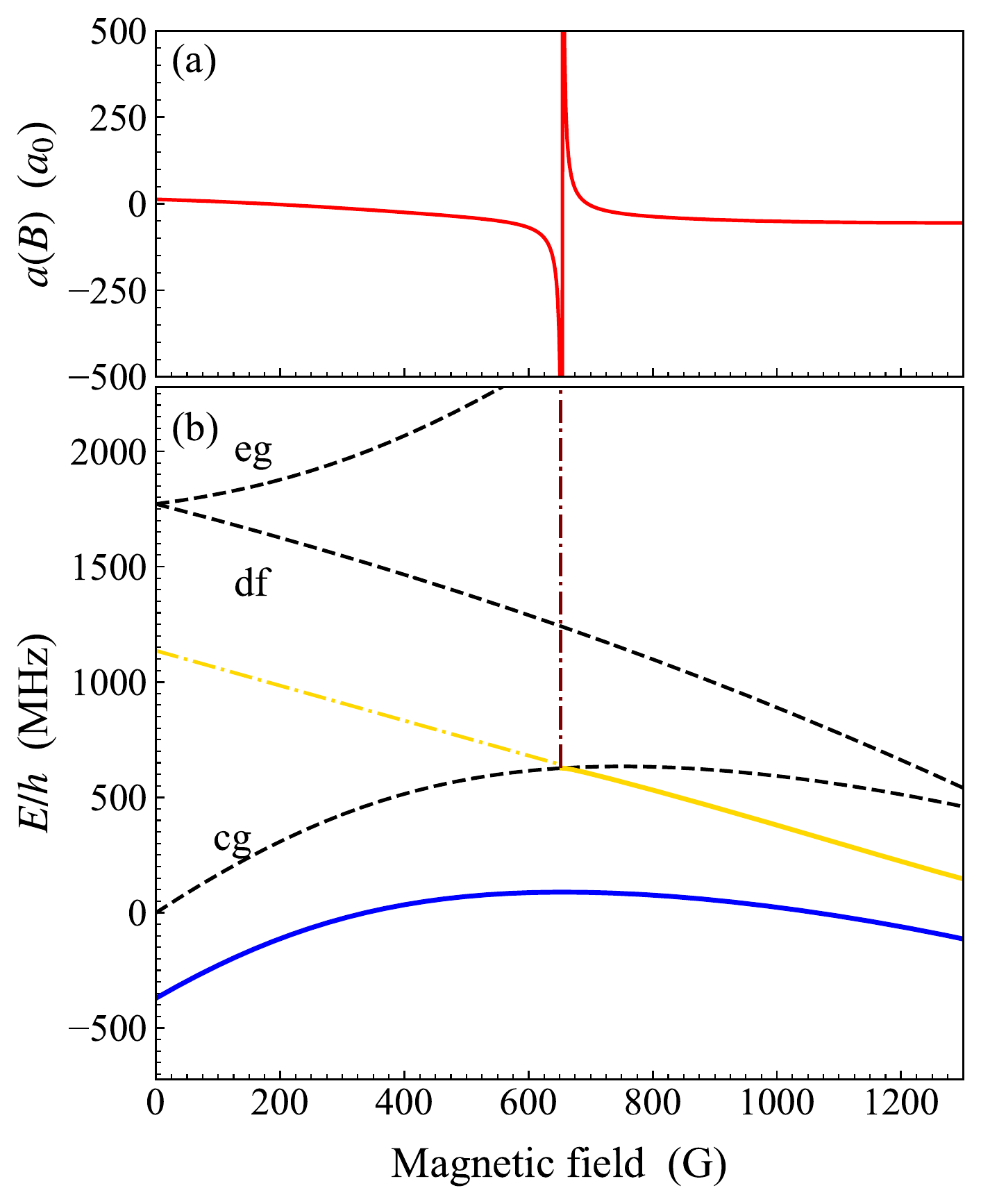}
    \caption{(a) The calculated s-wave scattering length in the cg channel as a function of magnetic field. (b) Energies of weakly bound s-wave molecular states with $M_F=-4$ (solid lines) and of nearby thresholds (dashed lines). The zero of energy is the zero-field energy of the cg and aa thresholds.
    The resonant state (yellow) is approximately parallel to the df threshold and there is another state (blue) roughly parallel to the cg threshold. The resonance position is marked by a vertical line extending up towards the scattering length plot. The dot-dash yellow line shows a linear extrapolation of the resonant state to zero field.}
    \label{fig:cg_states}
\end{figure}

Hood et al.\ \cite{Hood:NaCs:2020} have measured the position of an inelastic loss feature in the cg channel at 652.1(4)~G. This channel corresponds to $(f,m_f)=(1,-1)+(3,-3)$ at low field. They attributed this feature to an s-wave Feshbach resonance, and its position is designated $B_\textrm{res}^\textrm{cg}$.

The state that causes this resonance crosses downwards across the threshold with increasing magnetic field. It is bound at fields above the crossing, but is quasibound at fields below it, so cannot as simply be traced back to its origin at zero field with \textsc{bound}. Figure \ref{fig:cg_states} shows the bound states and atomic thresholds with $M_F=-4$ relevant to this resonance. A least-squares fit to the crossing state (solid yellow line) at fields above the crossing gives a gradient of $-0.76$~MHz/G and a zero-field intercept of $-5140$ MHz. The state is reasonably parallel to the df threshold with $(f,m_f)=(2,-2)+(3,-2)$, which has a gradient of about $-0.7$~MHz/G; we conclude that the state is mostly of df character. Calculation of the wave function at a field 80~G above the crossing confirms this, though there is developing coupling to the state in the cg channel (solid blue line) with increasing field. The state is bound by about $640$ MHz with respect to the df threshold, indicating that it lies in the top bin. Its overall triplet fraction is 60.6\%.

This state has a roughly similar triplet fraction and binding energy (with respect to the threshold that supports it) as the least-bound state in the ha channel. However, the interpretation of the position of the loss peak is somewhat uncertain. First, the resonance is quite broad, as seen in Fig.\ \ref{fig:cg_states}(a), with width $\Delta$ around 40~G. Secondly, Brooks et al.\ \cite{Brooks:2022} have shown that inelastic loss features for atom pairs in tweezers may be significantly shifted from the actual resonance position. We therefore conclude that the information on the interaction potential available from this feature is similar to, but less reliable than, that available from $E_{-1}^\textrm{ha}$; we therefore do not use $B_\textrm{res}^\textrm{cg}$ in fitting.

\subsubsection{Interaction shifts and derived scattering lengths}

Hood et al.\ \cite{Hood:NaCs:2020} have measured interaction shifts for spin-flip transitions of Na atoms (transition a$\leftrightarrow$h) and Cs atoms (transition a$\leftrightarrow$p) in tweezers. The shifts are defined as the difference in transition frequency between a tweezer containing one atom of each species and a tweezer containing a single atom. They are made up of shifts for individual pair states that depend on the scattering length for the particular pair of atomic states. However, modeling the shift for two different atoms in a non-spherical tweezer involves a complicated forwards calculation to take account of the anisotropy of the trap and the coupling between the relative and center-of-mass motions of the atoms \cite{Hood:NaCs:2020}.

Hood et al.\ used their measurement of $E_{-1}^\textrm{hp}$ to extract a triplet scattering length $a_\textrm{t}=30.4(6)\ a_0$. They used this to calculate the interaction shift for the hp state of Na+Cs, and hence to extract interaction shifts for the ha and ap states from the transition frequencies. They found an interaction shift of $-30.7$~kHz for the ha state, from which they inferred a large negative scattering length of $-693.8\ a_0$. From this they used MQDT to extract a singlet scattering length $a_\textrm{s}=428(9)\ a_0$.

The measurements of interaction shifts are principally sensitive the scattering length for the ha state. They contain information that is very similar to $E_{-1}^\textrm{ha}$, but is less precise and far less direct. We therefore do not use them in fitting.

\subsection{Fitting potential parameters} \label{sec:fitting}

The interaction potentials of Docenko et al.\ \cite{Docenko:2006} were fitted primarily to FT spectra, which accurately determine the deeper part of the potential but not the near-threshold part. Our goal is to adjust the potential curves to fit the ultracold observables described above, while retaining as much as possible their ability to reproduce the FT spectra. We therefore keep the two power series that represent the singlet and triplet potential wells fixed, with the coefficients obtained in Ref.\ \citenum{Docenko:2006}, and vary only the short-range and long-range extrapolations. As will be seen below, we found it necessary to make small changes in the long-range dispersion coefficients $C_6$ and $C_8$ of Eq.\ (\ref{eq:lr}), as well as to vary the parameters of the short-range extrapolations, $R_{{\rm SR},S}$ and $N_S$ of Eq.\ (\ref{eq:sr}).

There is no advantage in varying $R_{\textrm{LR},S}$, the point at which the mid-range power series (\ref{eq:mid}) is matched to the long-range exchange-dispersion potential (\ref{eq:lr}). As described above, continuity of the curves at $R_{\textrm{LR},S}$ is achieved by shifting the mid-range curves bodily using the constant terms $a_{0,S}$ in the power series. Any change in the dispersion coefficients $C_6$ and $C_8$ thus shifts the minima of both curves, and is directly reflected in the binding energy $E_{00}$ of the absolute ground state. The measured value of $E_{00}$ effectively provides a constraint that relates $C_8$ to $C_6$.

For a single potential curve $V(R)$ that varies as $-C_6/R^6$ at long range, the scattering length $a$ is approximately related to a phase integral $\Phi$ by \cite{Gribakin:1993}
\begin{equation}
a = \bar{a} \left[ 1 - \tan\left(\Phi - \frac{\pi}{8} \right) \right],
\end{equation}
where
\begin{equation}
\Phi = \int_{R_\textrm{in}}^\infty \left(2\mu[E_\textrm{thresh}-V(R)]/\hbar^2\right)^\frac{1}{2} dR
\end{equation}
and $R_\textrm{in}$ is the inner classical turning point at the threshold energy $E_\textrm{thresh}$. With the mid-range and long-range parts of the curve fixed by other observables, the only way to adjust $a$ is to vary the short-range potential in the region between $R_\textrm{in}$ and $R_\textrm{SR}$, where it is given by Eq.\ (\ref{eq:sr}). Since the relationship between $a$ and the binding energy $E_{-1}$ is only very weakly affected by the dispersion coefficients, the same applies to $E_{-1}$. These considerations apply independently to the singlet and triplet curves, so we have dropped the $S$ subscript here.

\begin{figure}[tbp]
 \centering
    \includegraphics[width=1\columnwidth]{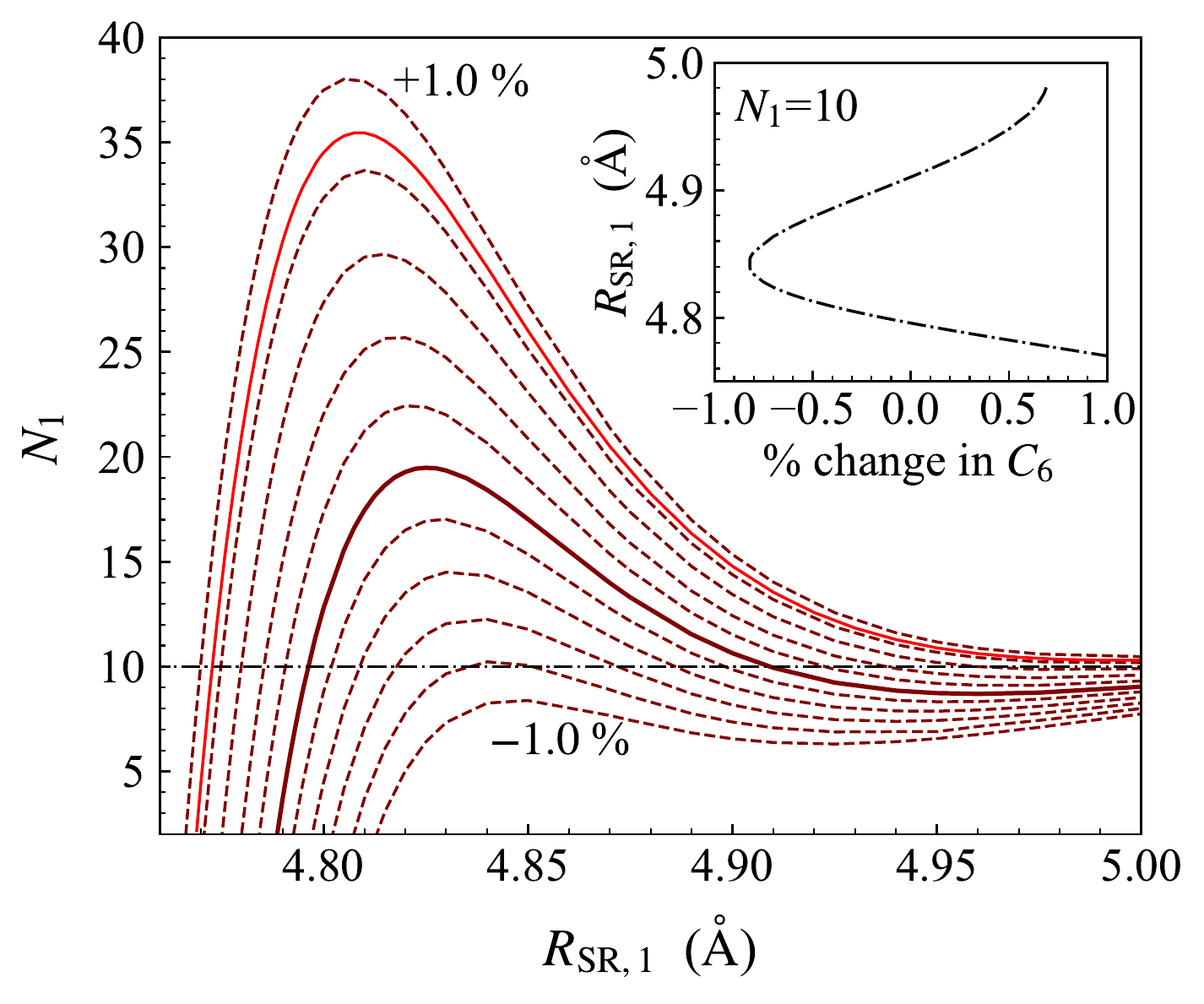}
    \caption{The relationship between the inverse power $N_1$ and the short-range matching point $R_\textrm{SR,1}$ required to reproduce the experimental binding energy $E_{-1}^\textrm{hp}$ of the least-bound triplet state of NaCs. The relationship is given for various values of the dispersion coefficient $C_6$, expressed as percentage differences from the theoretical value \cite{Derevianko:2001}. The solid brown line shows the value used in ref.\ \citenum{Docenko:2006} and the solid red line shows the final value of the present work. The inset shows the dependence of $R_\textrm{SR,1}$ on $C_6$ for the choice $N_1=10$.}
    \label{fig:triplet_fitting}
\end{figure}

If $A_\textrm{SR}$ and $B_\textrm{SR}$ are chosen to give continuity of the potential and its derivative at $R_\textrm{SR}$, the short-range extrapolation (\ref{eq:sr}) for each curve has free parameters $R_\textrm{SR}$ and $N$. The short-range power $N$ controls the hardness of the repulsive wall, and can substantially affect the extrapolation of the potential to energies above dissociation, which are important for higher-energy collisions. Nevertheless, in potentials fitted to FT spectra, $N$ has commonly been assigned an arbitrary fixed value, which has ranged from 3 for NaCs \cite{Docenko:2006} to 12 for K$_2$ \cite{Tiemann:2020}. A requirement to reproduce a particular value of $a$ or $E_{-1}$ is satisfied along a line in the space of $R_\textrm{SR}$ and $N$. However, because of the longer-range contribution to the phase integral $\Phi$, this line depends significantly on the values of $C_6$ and $C_8$.

We apply this approach first to the potential curve for the triplet state. As described above, the \textsc{field} package can automatically converge on the value of a potential parameter (here $R_\textrm{SR,1}$) required to reproduce a particular observable (here $E_{-1}^\textrm{hp}$). The resulting curves that relate $N_1$ and $R_\textrm{SR,1}$ are shown in Fig.\ \ref{fig:triplet_fitting}. The curves do depend on $C_6$ and the associated $C_8$, so are shown for values of $C_6$ that vary by up to $\pm1$\% from the theoretical value of ref.\ \citenum{Derevianko:2001}. As described below, $N_1$ will ultimately be chosen on physical grounds, and the inset of Fig.\ \ref{fig:triplet_fitting} shows how the required value of $R_\textrm{SR,1}$ depends on $C_6$ for the choice $N_1=10$.

\begin{figure}[tbp]
 \centering
    \includegraphics[width=1\columnwidth]{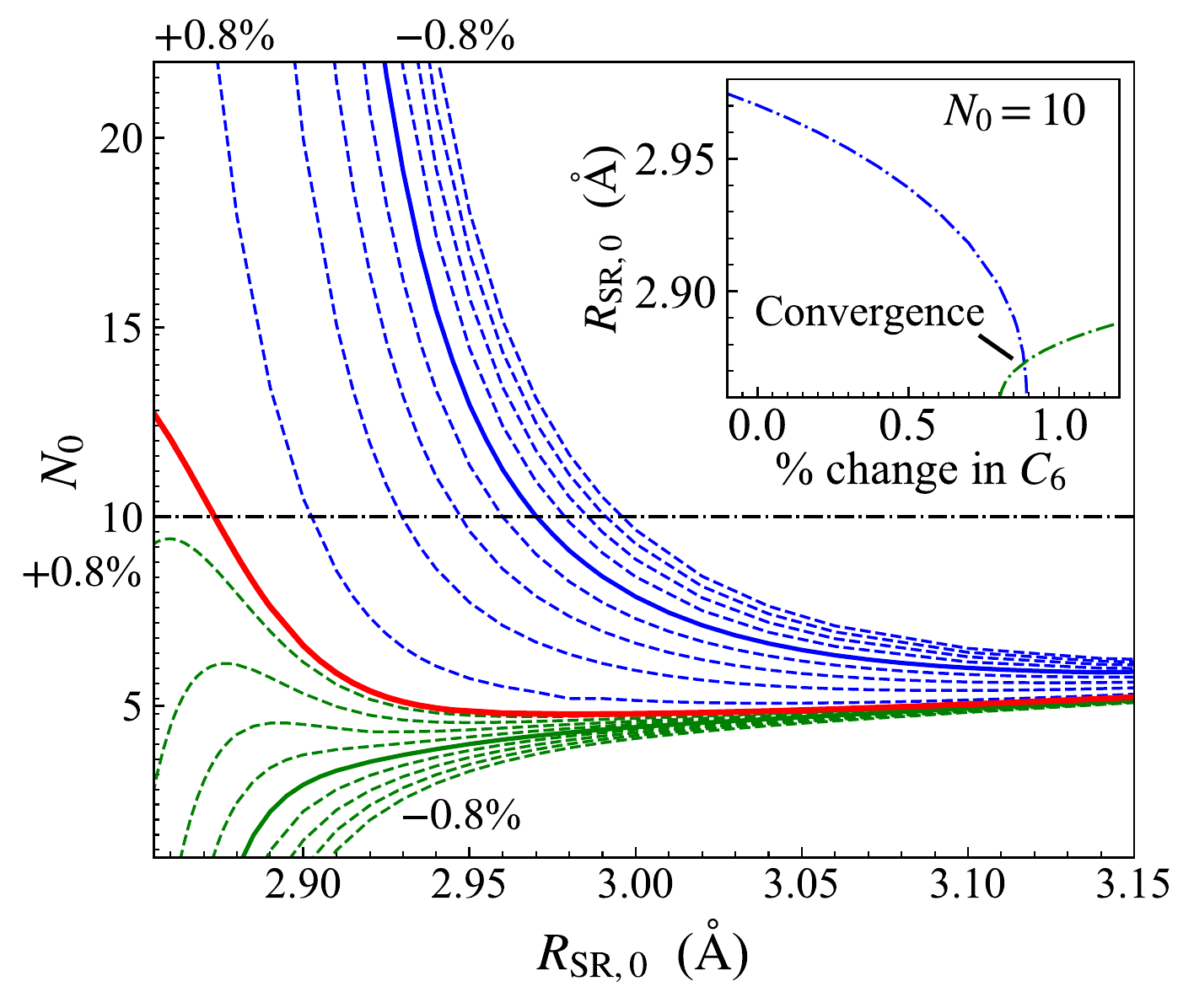}
    \caption{The relationship (green dashed lines) between the inverse power $N_0$ and the short-range matching point $R_\textrm{SR,0}$ required to reproduce the experimental binding energy $E_{-1}^\textrm{ha}$ of the least-bound state of NaCs in the ha channel. The relationship is given for various values of the dispersion coefficient $C_6$, expressed as percentage differences from the theoretical value \cite{Derevianko:2001}. The solid green curve shows the value used in ref.\ \citenum{Docenko:2006}. The blue lines show the analogous relationships required to reproduce the experimental position $B_\textrm{res}^\textrm{aa}$ of the s-wave resonance in the aa channel of Na+Cs. The solid red line is for the values of $C_6$ required to reproduce $E_{-1}^\textrm{ha}$ and $B_\textrm{res}^\textrm{aa}$ simultaneously. The inset shows the dependence of $R_\textrm{SR,1}$ on $C_6$ required to fit each observable for the choice $N_1=N_0=10$.}
    \label{fig:singlet_fitting}
\end{figure}

Once values are chosen for $C_6$, $C_8$, $N_1$ and $R_\textrm{SR,1}$, the triplet curve is fully defined. The same procedure may then be applied to vary the short-range part of the singlet curve to reproduce $E_{-1}^\textrm{ha}$. Since this state has multiple components as shown in Fig.\ \ref{fig:ha_wf}, this requires coupled-channel bound-state calculations, but it is nevertheless conceptually similar. The resulting relationship between $R_\textrm{SR,0}$ and $N_0$ is shown by the green lines in Fig.\ \ref{fig:singlet_fitting}, again for a range of values of $C_6$.

We initially carried out this procedure with the dispersion coefficient $C_6$ of ref.\ \citenum{Derevianko:2001}, as used in ref.\ \citenum{Docenko:2006}. This produced the relationship between $R_\textrm{SR,1}$ and $N_1$ shown by the solid brown line in Figs.\ \ref{fig:triplet_fitting} and between $R_\textrm{SR,0}$ and $N_0$ by the solid green line in Fig.\ \ref{fig:singlet_fitting}. It may be seen that, for the original value of $C_6$, there is no value of $R_\textrm{SR,0}$ that fits $E_{-1}^\textrm{ha}$ for $N_0 \gtrsim 5$. Furthermore, the resulting potential curves fail to reproduce $B_\textrm{res}^\textrm{aa}$, the position of the resonance near 864~G in the aa channel; they place it near 873~G. This is because they place the zero-field binding energy of the state that causes this resonance significantly too deep, about 2470~MHz below the $(f_\textrm{Na}=2,f_\textrm{Cs}=3)$ thresholds that supports it. As seen in Fig.\ \ref{fig:aa_wf}, this is still a long-range state, whose binding energy is controlled by the singlet and triplet scattering lengths and the dispersion coefficients. However, its wave function does not extend as far to long range as the least-bound states in Fig.\ \ref{fig:ha_wf}, so its binding energy is more sensitive to the dispersion coefficients than theirs. Since the relationship between $C_6$ and $C_8$ is determined by the binding energy of the absolute ground state, and the singlet and triplet scattering lengths are determined by $E_{-1}^\textrm{ha}$ and $E_{-1}^\textrm{hp}$, the only way to adjust $B_\textrm{res}^\textrm{aa}$ is by varying $C_6$ and $C_8$.

We therefore repeat the calculation of the relationship between $R_\textrm{SR,0}$ and $N_0$, but fitting to $B_\textrm{res}^\textrm{aa}$ instead of $E_{-1}^\textrm{ha}$. This produces the blue lines in Fig.\ \ref{fig:singlet_fitting}, again for a range of values of $C_6$. It may be seen that the lines fitted to $B_\textrm{res}^\textrm{aa}$ and to $E_{-1}^\textrm{ha}$ are incompatible unless $C_6$ is increased from its original value by approximately 0.9\%. The inset of Fig.\ \ref{fig:singlet_fitting} shows the values of $R_\textrm{SR,0}$ obtained from each of the two fits for the choice $N_0=N_1=10$. The requirement to fit both quantities produces a single value of $C_6$ (and the corresponding $C_8$ as required to reproduce $E_{00}$ as above).

These results led us to an iterative procedure for fitting the experimental observable. We (i) choose values for $N_0$, $N_1$ and $C_6$; (ii) vary $C_8$ to fit $E_{00}$; (iii) vary $R_\textrm{SR,1}$ to fit $E_{-1}^\textrm{hp}$; (iv) vary $R_\textrm{SR,0}$ to fit $E_{-1}^\textrm{ha}$; (v) evaluate $B_\textrm{res}^\textrm{aa}$, adjust $C_6$, and return to (ii). We repeat this cycle until convergence is achieved. This can be done for any reasonable values of $N_0$ and $N_1$, with results shown by the red line in Fig.\ \ref{fig:triplet_fitting} and by the red line in \ref{fig:singlet_fitting} for the choice $N_1=10$. Any potential along these lines reproduces the 4 observables $E_{00}$, $E_{-1}^\textrm{hp}$, $E_{-1}^\textrm{ha}$ and  $B_\textrm{res}^\textrm{aa}$, and they differ very little in their predictions for other observable quantities. For our final interaction potential, we choose $N_0=N_1=10$ to avoid the very soft repulsive wall of the triplet curve in ref.\ \citenum{Docenko:2006}.

\begin{table}[tbp]
\caption{\label{tab:parameter} Parameters of the fitted interaction potential, including the resulting singlet and triplet scattering lengths. Only quantities that are different from those of ref.\ \citenum{Docenko:2006} are listed. The derived parameters $A_{\textrm{SR}}$, $B_{\textrm{SR}}$ and $a_{0,S}$, which arise from the continuity constraints applied to $V(R)$ and $V'(R)$, are included for convenience in evaluating the potential curves. The rounded values of $A_{\textrm{SR}}$ correspond to the rounded values of $B_{\textrm{SR}}$, and differ slightly from the values obtained with the exact $B_{\textrm{SR}}$.}
\begin{center}
\renewcommand{\arraystretch}{1.0}
\begin{tabular}{ccc}
\hline
& Singlet & Triplet \\
\hline
$R_{\textrm{SR},S}$ (\AA) & 2.873\,240(6\,000)  & 4.772\,797(1\,600) \\
$N_S$ (\AA) & 10 &  10 \\
$A_{\textrm{SR},S}/hc$ (cm$^{-1}$) & $-3798.0168$ & $-420.536$ \\
$B_{\textrm{SR},S}/hc$ (cm$^{-1}$ \AA$^{10}$)& $1.309\,71 \times 10^8$ & $2.560\,41 \times 10^9$\\
$a_{0,S}/hc$ (cm$^{-1}$) & $-4954.229\,485$ & $-217.146\,766$  \\
 & & \\
$C_6/hc\ (10^7$ cm$^{-1}$ \AA$^6$) & \multicolumn{2}{c}{1.568\,975(400)} \\
$C_8/hc\ (10^8$ cm$^{-1}$ \AA$^8$) & \multicolumn{2}{c}{4.815\,171(5\,000)} \\
& & \\
$a_{\textrm{s}}$ or $a_{\textrm{t}}\ (a_0)$ & 433.05(65) & 30.55(22) \\
\hline

\end{tabular}
\end{center}
\end{table}

\begin{figure}[tbp]
 \centering
    \includegraphics[width=\columnwidth]{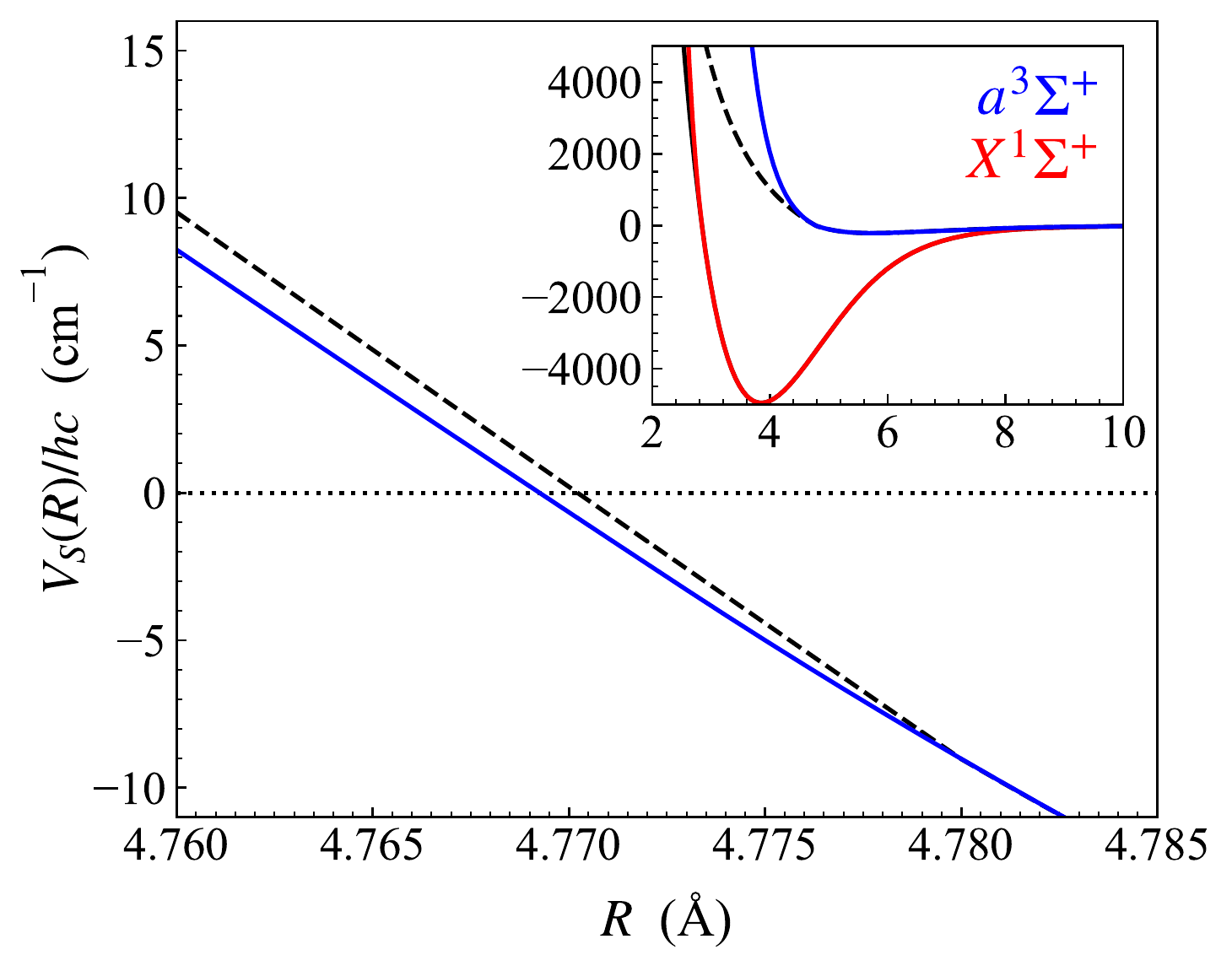}
    \caption{Comparison of the short-range region of the triplet curve of the present work (blue) with that of ref.\ \citenum{Docenko:2006} (dashed black). The labeled dot-dashed lines indicate $R_\textrm{SR,1}$ and the corresponding potential energies for the two curves. The derivative discontinuity in the potential curve of ref.\ \  is clearly visible. The inset shows the complete potential wells and the extrapolations onto the repulsive wall, including the singlet curve (red for the present work).}
    \label{fig:differences}
\end{figure}

It would have been possible to obtain the same final potential by a ``blind" minimization procedure, but it conveys important insights to understand the interplay between parameters and the lines in parameter space that are capable of fitting each observable.

The parameters that differ from those of ref.\ \citenum{Docenko:2006} are given in Table \ref{tab:parameter}, together with the resulting singlet and triplet scattering lengths. Compared to ref.\ \citenum{Docenko:2006}, $R_\textrm{SR,0}$ and $R_\textrm{SR,1}$ have changed by 0.03 \AA\ and $-0.0072$ \AA, respectively; $N_S$ has been fixed at a more physically reasonable value of 10 for both states, compared to its original value of 3; $C_6$ has increased by 0.9\%; in atomic units it is 3257(1) $E_\textrm{h} a_0^6$, compared with 3227(18) $E_\textrm{h} a_0^6$ from ref.\ \citenum{Derevianko:2001}; and $C_8$ has decreased by 3\% from the fitted value of ref.\ \citenum{Docenko:2006}, but our fitted value corresponds to $C_8=3.568(4) \times 10^5\ E_\textrm{h} a_0^8$, which is closer to the theoretical value of $C_8=3.62(12) \times 10^5\ E_\textrm{h} a_0^8$ \cite{Porsev:2003} and well within its uncertainty.

Key differences between our potential curves and those of ref.\ \citenum{Docenko:2006} are shown in Fig.\ \ref{fig:differences}. The derivative discontinuity in the triplet potential of ref.\ \citenum{Docenko:2006} is clearly visible at 4.78 \AA. The present triplet potential continues smoothly through $R_\textrm{SR,1}$, so has a zero-energy turning point at slightly shorter range, 4.7693 \AA, compared to 4.7702~\AA\ for the potential of ref.\ \citenum{Docenko:2006}. The effect of the larger values of $N_0$ and $N_1$ is seen in the steeper short-range repulsive walls shown in the inset.

\subsubsection{Uncertainties in fitted parameters}

The interaction potential obtained here is obtained by fitting four potential parameters to four experimental quantities. The 4-parameter space is actually a subspace of a much larger space, of approximately 50 parameters, that were fitted to FT spectra in ref.\ \citenum{Docenko:2006}. ref.\ \citenum{Docenko:2006} itself gave no uncertainties for the fitted parameters or estimates of the correlations between them. It is therefore not appropriate or practical to use error estimates based on deviations between observed and calculated properties. We can nevertheless make estimates of errors based on the derivatives of the calculated observables with respect to potential parameters, as described in Appendix \ref{app:uncer}, and these are included in Table \ref{tab:parameter}.

\subsection{Predictions of the fitted potential} \label{sec:predict}

\subsubsection{Scattering lengths}

The singlet and triplet scattering lengths given in Table \ref{tab:parameter} are within the uncertainties of those obtained by Hood et al.\ \cite{Hood:NaCs:2020}, $a_\textrm{s}=428(9)\ a_0$ and $a_\textrm{t}=30.4(6)\ a_0$. Their value of $a_\textrm{t}$ was obtained from $E_{-1}^\textrm{hp}$, so is of similar accuracy to ours, though ours is shifted slightly because we have determined improved values of the dispersion coefficients. Their value of $a_\textrm{s}$ was obtained by combining $a_\textrm{t}$ with measurements of interaction shifts, as described above. Our value of $a_\textrm{s}$ is considerably more precise, both because of the greater precision of $E_{-1}^\textrm{ha}$ compared to the interaction shifts and because of the use of full coupled-channel calculations.

Hood et al.\ also gave the scattering length for the ha channel as $-693\ a_0$, without an error estimate. This quantity is important because the large negative value enhances the intensity of photoassociation transitions originating from atoms in the ha state \cite{Zhang:NaCs:2020}. Our interaction potential gives an even larger negative value of $-860(2)\ a_0$. The value of ref.\ \citenum{Hood:NaCs:2020} arose fairly directly from their measurements of interaction shifts, which are dominated by the ha channel. Our value is principally based on the more reliable and precise measurement of $E_{-1}^\textrm{ha}$, so is expected to be more accurate.

\subsubsection{Bound states with $L=0$}

\begin{figure}[tbp]
 \centering
    \includegraphics[width=\columnwidth]{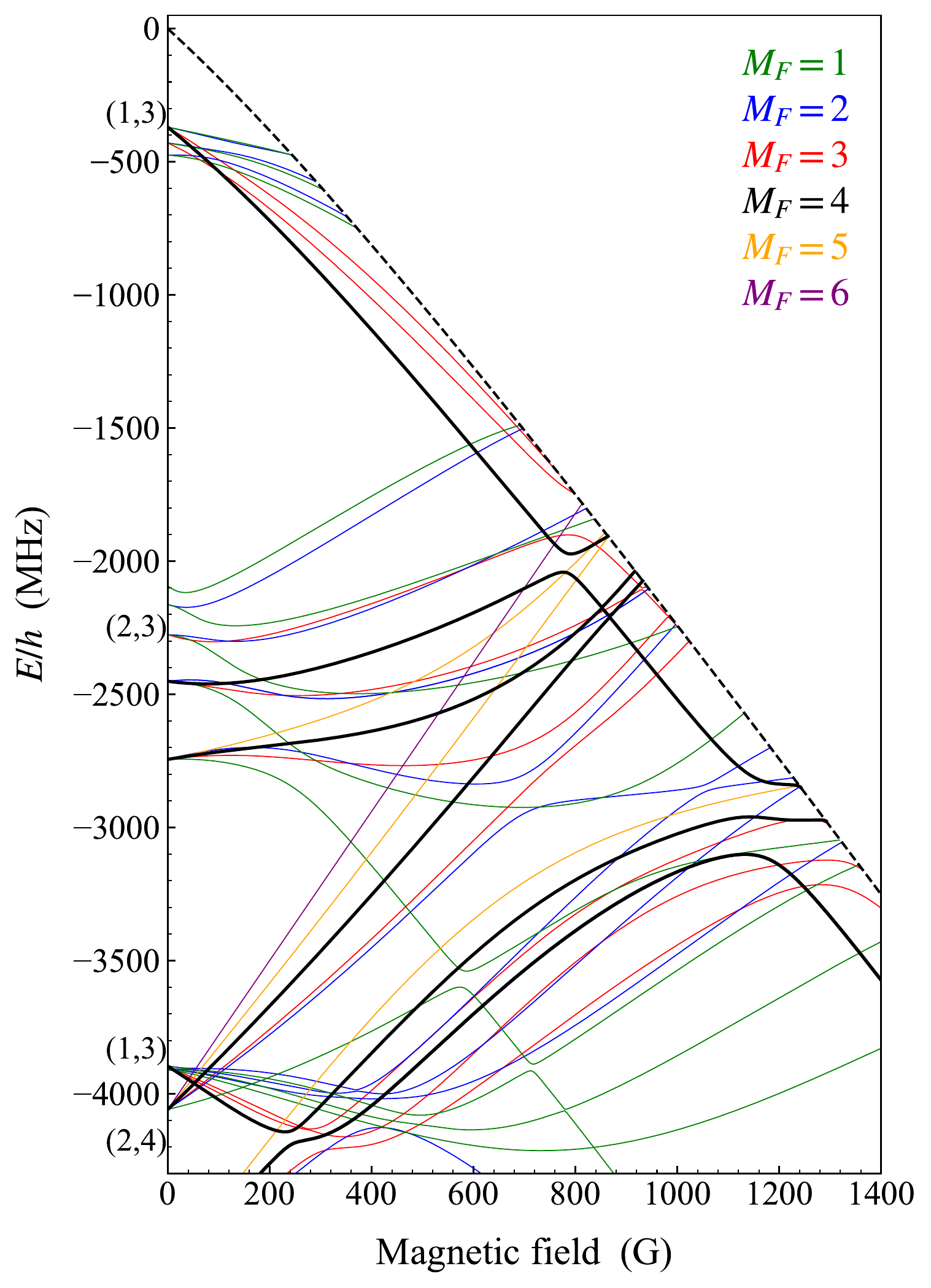}
    \caption{Weakly bound states of NaCs with $L=0$ below the aa threshold as a function of magnetic field. The aa threshold is shown as a dashed black line. States with $M_F=4$ that can cause s-wave Feshbach resonances are shown as solid black lines; other values of $M_F$ are color-coded as shown in the legend. Only states with $M_F$ from 1 to 6 are shown. The zero of energy is the threshold energy at zero field, which lies 6278.1~MHz below the hyperfine centroid.}
    \label{fig:bound}
\end{figure}

Figure \ref{fig:bound} shows the energies of bound states of NaCs below the lowest (aa) threshold, as a function of magnetic field. All states with $M_F$ between 1 and 6 are included (but not states with $M_F$ from $-6$ to 0). The calculation uses a basis set with $L_\textrm{max}=0$, so only states with $L=0$ are shown. At zero field, the states can be grouped according to their hyperfine characters. The uppermost group, with zero-field binding energies from $350$ to $500$ MHz, are $n=-1$ states with character ($f_{\textrm{Na}},f_{\textrm{Cs}}$) = (1,3). The next group, from 2000 to 2800 MHz, are $n=-2$ states with character (2,3). The group near 3900~MHz have character (1,3) but with $n=-2$. Finally, the deepest group shown, which starts slightly deeper than 4000 MHz and extends off the bottom of the plot, are $n=-3$ states with character (2,4).

For each group, $f_{\textrm{Na}}$ couples to $f_{\textrm{Cs}}$ to give a resultant $F$, which is a good quantum number at zero field. The allowed values of $F$ run from $f_{\textrm{Cs}}-f_{\textrm{Na}}$ to $f_{\textrm{Cs}}+f_{\textrm{Na}}$ in steps of 1. In a magnetic field, each state splits into components with different $M_F$ (though not all possible values of $M_F$ are shown). The value of $F$ for a zero-field state can therefore be inferred from the largest $M_F$ present. $M_F$ is a good quantum number when $L_\textrm{max}=0$, but at moderate fields (between 30 and 500~G) states of the same $M_F$ but different $F$ approach one another and mix; above these fields, $m_{f,{\textrm{Na}}}$ and $m_{f,{\textrm{Ca}}}$ are better quantum numbers than $F$.

\subsubsection{Resonances in s-wave scattering} \label{sec:res-s}

\begin{figure}[tbp]
 \centering
    \includegraphics[width=\columnwidth]{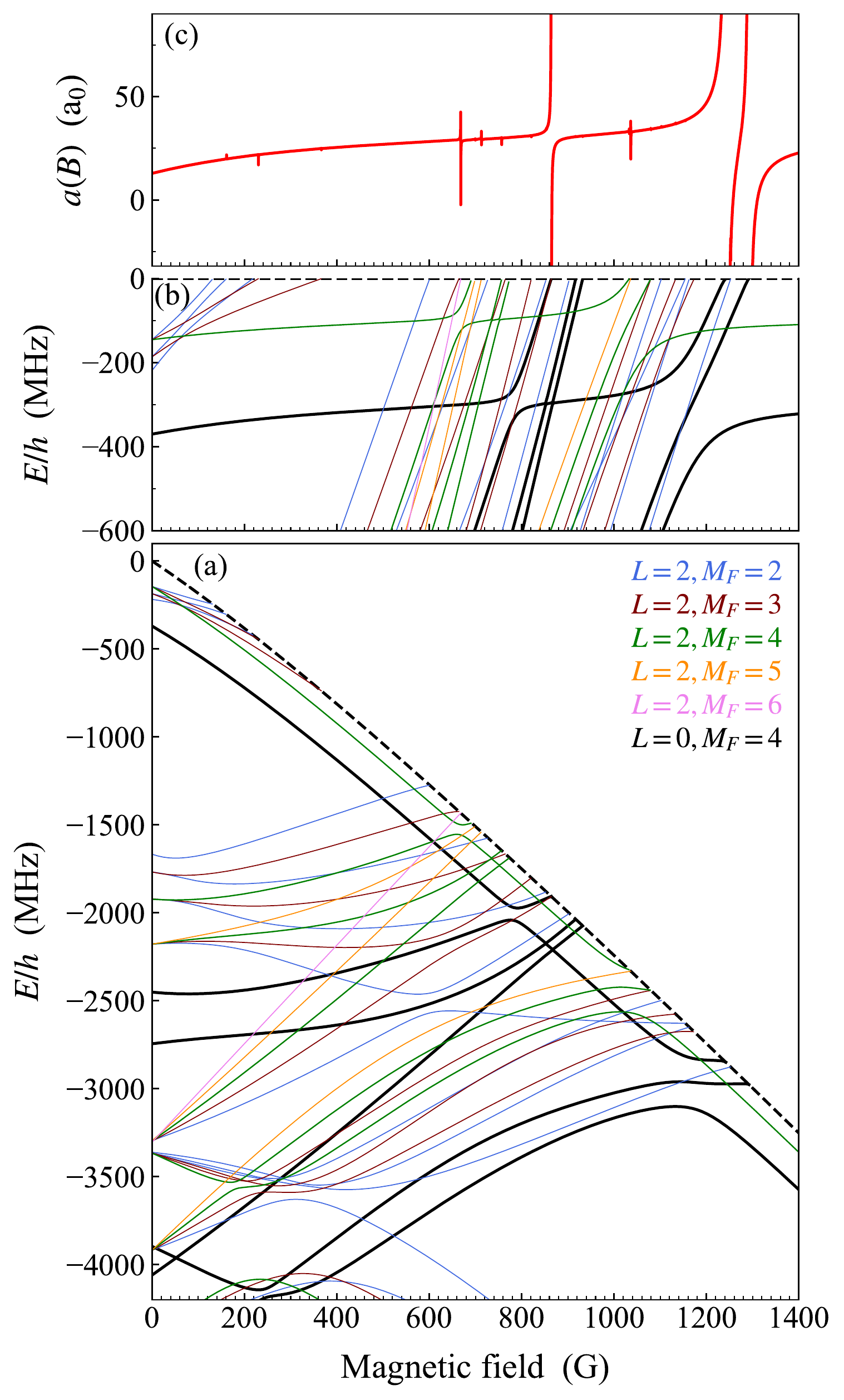}
    \caption{(a) Weakly bound states of NaCs with $M_\textrm{tot}=4$ and $L=0$ or 2 below the aa threshold as a function of magnetic field. The aa threshold is shown as a dashed black line. States with $L=0$ and $M_F=4$ that can cause s-wave Feshbach resonances are shown as solid black lines; states with $L=2$ that can cause d-wave resonances are color-coded according to $M_F$ as shown in the legend. The zero of energy is the threshold energy at zero field. (b) Expanded view of (a), with energies shown as binding energies with respect to the aa threshold. (c) s-wave scattering length at the aa threshold, showing resonances where bound states cross threshold. Some of the resonances that exist are too narrow to see on the 0.2~G grid used for the calculation of the scattering length.}
    \label{fig:d-wave}
\end{figure}

It is important to distinguish between $L_\text{in}$ for the incoming wave and $L$ for a bound state. The widest resonances in s-wave scattering ($L_\text{in}=0$) are due to s-wave bound states (with $L=0$), and are referred to as s-wave resonances. Since $M_\textrm{tot} = M_F + M_L$ is conserved and is 4 for an incoming s wave at the aa threshold, bound states with $L=0$ can cause resonances at this threshold only if they have $M_F=4$. These states are shown as solid black lines in Fig.\ \ref{fig:bound}.

Bound states with even $L>0$ can also cause Feshbach resonances in s-wave scattering, which are usually narrower. The widest of these are d-wave resonances, due to d-wave states (with $L=2$). In this case $M_L$ can take values from $-2$ to 2, so d-wave states with $M_F$ = 2 to 6 can have $M_\textrm{tot}=4$ and cause resonances in s-wave scattering at the aa threshold.

Figure \ref{fig:d-wave}(a) shows all states with $M_\textrm{tot}=4$ that lie close to the aa threshold, as a function of magnetic field. This calculation uses a basis set with $L_\textrm{max}=2$, so includes states with both $L=0$ and 2. States with $L=0$ and $M_F=4$ are again shown in black, whereas states with $L=2$ are color-coded according to $M_F$. To allow this labeling, the small couplings off-diagonal in $M_F$ are neglected in the bound-state calculations (but not in the corresponding scattering calculations). The pattern of zero-field states for each hyperfine group is similar in structure to Fig.\ \ref{fig:bound}, but the states with $L=2$ are shifted upwards by a rotational energy. Figure \ref{fig:d-wave}(b) shows an expanded view of the bound states, plotted as energies below the aa threshold, and Figure \ref{fig:d-wave}(c) shows the resulting s-wave scattering length. A resonance occurs at every field where a state with $M_\textrm{tot}=4$ crosses threshold, but some of them are too narrow to be visible on the grid of magnetic fields used for Fig.\ \ref{fig:d-wave}(c). Nevertheless, all of them can be characterized in scattering calculations, using the methods of ref.\ \citenum{Frye:resonance:2017}, to give values for $B_\textrm{res}$, $\Delta$ and $a_\textrm{bg}$ from Eq.\ \ref{eq:res}.

\begin{table}[tbp]
\caption{\label{tab:resonances} Feshbach resonances with widths greater than $10^{-4}$~G in s-wave and p-wave scattering at the aa threshold. The p-wave calculations are for $M_\textrm{tot}=4$ only.}
\begin{center}
\begin{tabular}{rcccc}
\hline
\multicolumn{5}{c}{Resonances in s-wave scattering (34 total)} \\
\hline
$B_\textrm{res}$ (G) & $\Delta$ (G) & $a_\textrm{bg}\ (a_0)$ & $L$ & $M_F$  \\
 161.23 & 0.0007 & 19.8    & 2 & 2  \\
 218.30 & 0.0002 & 21.6    & 2 & 2  \\
 230.24 & 0.0007 & 21.9    & 2 & 3  \\
 366.36 & 0.0010 & 24.8    & 2 & 3  \\
 668.14 & 0.066  & 28.9    & 2 & 6  \\
 699.69 & 0.0012 & 29.2    & 2 & 5  \\
 712.89 & 0.011  & 29.4    & 2 & 5  \\
 756.80 & 0.0016 & 29.9    & 2 & 4  \\
 773.90 & 0.0002 & 30.2    & 2 & 4  \\
 853.50 & 0.0008 & 34.2    & 2 & 2  \\
 864.13 & 1.27   & 30.7    & 0 & 4  \\
 864.42 &$-0.0001$& $-105$ & 2 & 3  \\
 917.07 & 0.0003 & 30.6    & 0 & 4  \\
 932.20 & 0.0003 & 30.9    & 0 & 4  \\
1032.90 & 0.0035 & 33.2    & 2 & 4  \\
1036.15 & 0.022  & 33.0    & 2 & 5  \\
1080.00 & 0.001  & 34.3    & 2 & 3  \\
1133.52 & 0.0005 & 36.9    & 2 & 3  \\
1243.02 & 14.4   & 40.2    & 0 & 4  \\
1252.53 &$-0.026$&$-22.6$  & 2 & 2  \\
1292.57 & 17.7   & 20.5    & 0 & 4  \\
\hline
\hline
\multicolumn{5}{c}{Resonances in p-wave scattering (17 total)} \\
\hline
$B_\textrm{res}$ (G) & $\Delta$ (G) & $v_\textrm{bg}\ (10^7\ a_0^3)$ & $L$ & $M_F$  \\
\hline
 805.41 & 0.021  & $-1.50$ & 1 & 4  \\
 806.80 & 0.0083 & $-1.53$ & 1 & 5  \\
1173.87 & 0.42   & $-1.50$ & 1 & 4  \\
1216.83 & 0.0067 & $-1.47$ & 1 & 3  \\
1222.78 & 0.21   & $-1.53$ & 1 & 4  \\
\hline
\end{tabular}
\end{center}
\end{table}

Table \ref{tab:resonances} gives the parameters of all s-wave and d-wave resonances with $\Delta>10^{-4}$~G, together with quantum numbers for the states that cause them. It may be noted that the s-wave resonance near 864~G, which appeared at 864.11~G in a calculation with $L_\textrm{max}=0$, is shifted to 864.13~G in the calculation with $L_\textrm{max}=2$. This demonstrates the small effect of basis functions with $L=2$ on s-wave properties, and justifies the use of $L_\textrm{max}=0$ in fitting.

Zhang et al.\ \cite{Zhang:NaCs:2020} observed a weak d-wave Feshbach resonance at 864.5 G, on the shoulder of the s-wave resonance at 864.11~G. The bound state responsible for this is visible in Fig.\ \ref{fig:d-wave}(a), and crosses threshold at 864.42~G, causing a resonance of width $\Delta=-10^{-4}$~G. It is an impressive demonstration of the quality of our interaction potential that it can reproduce the position of this resonance to within 0.1~G and identify the bound state responsible: it is a state with $L=2$, $M_F=3$ (brown in Fig.\ \ref{fig:d-wave}), involving a pair of states originating from ($f_{\textrm{Na}},f_{\textrm{Cs}},F$) = (2,3,5) and (2,4,6) that experience an avoided crossing around 700~G.

\subsubsection{Resonances in p-wave scattering} \label{sec:res-p}

\begin{figure}[tbp]
 \centering
    \includegraphics[width=\columnwidth]{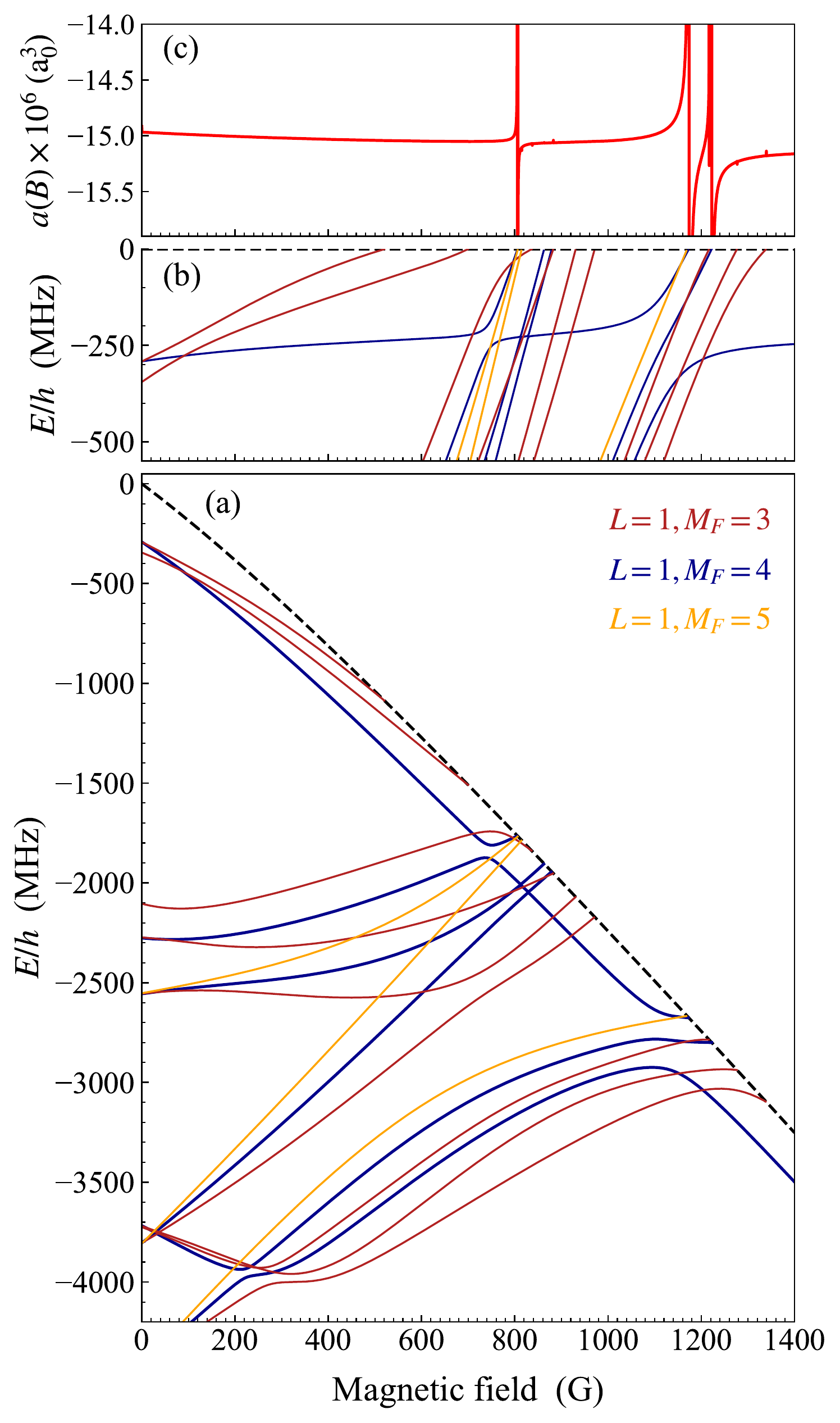}
    \caption{(a) Weakly bound p-wave states of NaCs, with $M_\textrm{tot}=4$ and $L=1$, below the aa threshold as a function of magnetic field. The aa threshold is shown as a dashed black line. Only states with $M_\textrm{tot}=4$ are shown. The states are color-coded according to $M_F$ as shown in the legend. The zero of energy is the threshold energy at zero field. (b) Expanded view of (a), with energies shown as binding energies with respect to the aa threshold. (c) p-wave scattering volume at the aa threshold, calculated at a collision energy of $2\ \mu\textrm{K}\times k_\textrm{B}$. Some of the resonances that exist are too narrow to see on the 0.2~G grid used for the calculation of scattering length.}
    \label{fig:p-wave}
\end{figure}

Resonances can also occur in p-wave scattering ($L_\textrm{in}=1$), due to either p-wave states (with $L=1$) or states with higher odd $L$. In the gas phase such resonances are usually observed only at relatively high temperatures (several $\mu$K), but in optical tweezers it is possible to enhance them selectively by promoting one atom to a motionally excited state. Zhang et al.\ \cite{Zhang:NaCs:2020} observed a group of p-wave resonances around 807~G for Na+Cs, with complicated structure, and used them to produce a single p-wave molecule in the tweezer.

For p-wave scattering, $M_{L,\textrm{in}}$ can be $-1$, 0 or $-1$ and $M_\textrm{tot}=M_{F,\textrm{in}}+M_{L,\textrm{in}}$. Thus, even at the aa threshold, $M_\textrm{tot}$ can be 3, 4 or 5. If the resonant state has $L=1$, $M_L$ can be $-1$, 0 or $-1$ too.
For each of the three values of $M_\textrm{tot}$, p-wave resonances arise from bound states with $M_F = M_\textrm{tot}$ and $M_\textrm{tot}\pm1$. Figure \ref{fig:p-wave}(c) shows the p-wave bound states below the aa threshold and the corresponding scattering volume $v$, but only for the case $M_{\textrm{tot}}=4$. The bound states show considerable similarities to the s-wave and p-wave ones in Figs.\ \ref{fig:bound} and \ref{fig:d-wave}. Figure \ref{fig:p-wave}(c) shows that s-wave and p-wave states share several similarities, but with shifts due to the different rotational energy in each case. The positions, widths and assignments of the widest resulting resonances are given in Table \ref{tab:resonances}, but it must be remembered that this is for only one of the three possible values of $M_\textrm{tot}$ for p-wave scattering at the aa threshold. Figure \ref{fig:p-wave} and Table \ref{tab:resonances} show that the group of resonances observed near 807~G \cite{Zhang:NaCs:2020} are mainly the p-wave analogs of the s-wave resonance near 864~G.

\subsubsection{Resonance in cg channel}

As described above, Hood et al.\ \cite{Hood:NaCs:2020} measured the position of an inelastic loss feature in the cg channel at 652.1(4)~G. Our fitted potential produces a resonance at 654.3~G. However, its width is $\Delta = 43$~G, so the difference between the resonance position and the observed loss peak is only 5\% of the width. The calculated background scattering length is $-41\ a_0$.

\section{Conclusions} \label{sec:conc}

We have used measurements on ultracold scattering and spectroscopy in optical tweezers \cite{Liu:NaCs:2019, Zhang:NaCs:2020, Hood:NaCs:2020, Yu:NaCs:2021, Cairncross:2021}, combined with previous work using Fourier-transform spectroscopy \cite{Docenko:2006}, to determine improved potential curves for the singlet and triplet states of NaCs. We have used coupled-channel calculations based on these curves to characterize the weakly bound states involved and to make predictions for additional bound states and Feshbach resonances.

Each measurement of a spectroscopic transition or resonance position is sensitive to the properties of one or two specific bound states of the molecule. These properties are in turn sensitive to particular features of the interaction potentials. Our work has produced important insights into these relationships, and the ways that combinations of measurements can be used to determine features of the potential curves.

For NaCs, as for many other diatomic molecules, the mid-range parts of the potential curves had previously been accurately determined from spectroscopy at relatively high temperatures. For NaCs, this mid-range part extends from just outside the inner turning point at the dissociation energy to 10.2~\AA, and is expressed as a power-series expansion for each of the singlet and triplet curves \cite{Docenko:2006}. Our approach is to change the mid-range part by as little as possible, to retain its ability to fit the higher-temperature spectra. We thus retain the mid-range expansion unchanged, and adjust only the extrapolations to long and short range. This gives sufficient flexibility to reproduce the ultracold observables.

The binding energy of the least-bound (top) bound state in a particular scattering channel, $E_{-1}$, is closely related to the scattering length $a$ for that channel. The relationship between $E_{-1}$ and $a$ depends on the dispersion coefficients for the long-range interaction, particularly $C_6$, but only weakly. Since the dispersion coefficients are often known fairly accurately from independent theory \cite{Derevianko:2001}, $E_{-1}$ is a good surrogate for $a$. If it can be measured for two channels that represent significantly different mixtures of singlet and triplet states, the singlet and triplet scattering lengths $a_\textrm{s}$ and $a_\textrm{t}$ can be disentangled. This is the case for NaCs, where $E_{-1}$ has been measured both for a spin-stretched channel that is pure triplet in character \cite{Liu:NaCs:2019, Hood:NaCs:2020} and for the ha channel \cite{Yu:NaCs:2021}, which has about 50\% singlet character. Since the mid-range part of the potential is held fixed to reproduce the higher-temperature spectra, and the dispersion coefficients have only limited influence, the two values of $E_{-1}$ determine the short-range parts of the singlet and triplet curves.

Magnetic Feshbach resonances exist where a weakly bound molecular state crosses a scattering threshold as a function of magnetic field. These states are often supported by thresholds in which one or both atoms are in excited hyperfine states. States that cause resonances at the lowest threshold are thus often bound by considerably more than the least-bound state. In NaCs, the state that causes the resonance observed in the lowest channel \cite{Zhang:NaCs:2020} is bound by more than 4~GHz with respect to the threshold that mostly supports it. Because of this, it is much more sensitive to the dispersion coefficients than the least-bound states. The requirement to reproduce this resonance position as well as the least-bound states places a strong constraint on the dispersion coefficients, particularly $C_6$.

In potential curves from higher-temperature spectroscopy, the dissociation energy (and thus the absolute binding energies of all the deeply bound states) is usually obtained from extrapolation, rather than measured directly. However, Raman transfer of ultracold molecules to a deeply bound state provides a direct measurement of its absolute binding energy. If the mid-range part of the potential is held fixed to reproduce the higher-temperature spectra, this provides a second (and different) constraint on the dispersion coefficients. Satisfying this along with the constraint from the resonance position allows $C_6$ and $C_8$ to be disentangled.

There is an important general insight here. The spectroscopy of ultracold molecules often provides measurement of the energies of the least-bound molecular states supported by one or more thresholds. Measurements of tunable Feshbach resonances are often sensitive to somewhat deeper states, with binding energies in the GHz range. When such measurements are combined, they can provide very precise values for dispersion coefficients. The same principle applies when different Feshbach resonances provide implicit information on two or more states with substantially different binding energies with respect to the thresholds that support them.

For NaCs, we find that the different ultracold observables can be fitted simultaneously only if $C_6$ is increased by about 0.9\% from the theoretical value. Our fitted value corresponds to 3256(1) $E_\textrm{h} a_0^6$, compared to 3227(18) $E_\textrm{h} a_0^6$ from ref.\ \citenum{Derevianko:2001}. Our fitted value $C_8=3.568(4) \times 10^5\ E_\textrm{h} a_0^8$ is well within the error bounds of the value of ref.\ \citenum{Porsev:2003}.

Accurately fitted interaction potentials are key to progress in ultracold scattering and spectroscopy. They provide predictions of new experimental observables, which are often crucial in designing experiments and locating new spectroscopic lines. They also provide calculated scattering lengths, as a function of magnetic field, which are unavailable from other sources. These are often crucial in experiments that need precise control of the scattering length, such as those exploring Efimov physics or quantum phase behavior.

\bigskip\noindent{\bf Data availability statement} \label{sec:data}

The data underlying this study are openly available from Durham Research Online at [DOI to be provided].

\begin{acknowledgments}
We are grateful to Matthew Frye, Ruth Le Sueur and Kang-Kuen Ni for valuable discussions.
This work was supported by the U.K. Engineering and Physical Sciences Research Council (EPSRC)
Grant No.\ EP/P01058X/1.
\end{acknowledgments}

\appendix
\section{Magnetic dipole interaction and second-order spin-orbit coupling} \label{app:SO}

At long range, the coupling $\hat V^{\rm d}(R)$ of Eq.\ \ref{eq:V-hat} has a simple magnetic dipole-dipole form that varies as $1/R^3$~\cite{Stoof:1988, Moerdijk:1995}. However, for heavy atoms such as Cs, second-order spin-orbit coupling provides an additional contribution that has the same tensor form as the dipole-dipole term and dominates at short range \cite{Mies:1996, Kotochigova:2000}. In the present work, $\hat V^{\rm d}(R)$ is written
\begin{equation}
\label{eq:Vd} \hat V^{\rm d}(R) = \lambda(R) \left ( \hat s_1\cdot
\hat s_2 -3 (\hat s_1 \cdot \vec e_R)(\hat s_2 \cdot \vec e_R)
\right ) \,,
\end{equation}
where $\vec e_R$ is a unit vector along the internuclear axis and $\lambda$ is an $R$-dependent coupling constant. This term couples the electron spins of Na and Cs atoms to the molecular axis.

The second-order spin-orbit splitting is not known for NaCs. However, it contributes only when $L_\textrm{max}>0$ and makes only very small contributions for s-wave states and resonances due to them. We model it here using the functional form used for RbCs \cite{Takekoshi:RbCs:2012},
\begin{align}
\label{eq:lambda}
\lambda(R) =& E_{\rm h} \alpha^2 \bigg[
A_{\rm 2SO}^{\rm short} \exp\left(-\beta_{\rm 2SO}^{\rm short}(R/a_0)\right)
\nonumber\\
+& A_{\rm 2SO}^{\rm long}
\exp\left(-\beta_{\rm 2SO}^{\rm long}(R/a_0)\right)
+  \frac{g_{S,\textrm{Na}} g_{S,\textrm{Cs}}}{4(R/a_0)^3}\bigg],
\end{align}
where $E_\textrm{h}$ is the Hartree energy and $\alpha\approx 1/137$ is the atomic fine-structure constant. To account for the smaller size of Na compared to Rb, we adjust the values of $A_{\rm 2SO}^{\rm short}$ and $A_{\rm 2SO}^{\rm short}$ for RbCs to shift the second-order spin-orbit contribution to short range by 0.757 $a_0$. This gives parameters $A_{\rm 2SO}^{\rm short} = -27.8$, $A_{\rm 2SO}^{\rm long} = -0.027$, $\beta_{\rm 2SO}^{\rm short} = 0.80$ and $\beta_{\rm 2SO}^{\rm long} = 0.28$ for NaCs. Future experiments may allow determination of these parameters, but changing them would have little effect on the singlet and triplet curves obtained here (though it might have a significant effect on the widths of predicted d-wave resonances).

\section{Coupled-channel methods} \label{app:cc}

We expand the total wave function of the molecule or colliding pair of atoms in a coupled-channel representation,
\begin{equation} \Psi(R,\xi)
=R^{-1}\sum_j\Phi_j(\xi)\psi_{j}(R). \label{eqexp}
\end{equation}
Here $\xi$ represents all coordinates of the pair except the internuclear distance $R$. The functions $\Phi_j(\xi)$ form a complete orthonormal basis set for motion in the coordinates $\xi$ and the factor $R^{-1}$ serves to simplify the action of the radial kinetic energy operator. The component of the wave function in each {\em channel} $j$ is described by $\psi_{j}(R)$, and these are the functions shown in Figs.\ \ref{fig:ha_wf} and \ref{fig:aa_wf}. The expansion (\ref{eqexp}) is substituted into the total Schr\"odinger equation, and the result is projected onto a basis function $\Phi_i(\xi)$. The resulting coupled differential equations for the functions $\psi_{i}(R)$ are
\begin{equation}
\frac{\d^2\psi_{i}}{\d R^2}
=\sum_j\left[W_{ij}(R)-{\cal E}\delta_{ij}\right]\psi_{j}(R), \label{eq:se-invlen}
\end{equation}
where $\delta_{ij}$ is the Kronecker delta, ${\cal E}=2\mu E/\hbar^2$, $E$ is
the total energy, and
\begin{align}
W_{ij}(R)&=\frac{2\mu}{\hbar^2}\int\Phi_i^*(\xi) [\hbar^2 \hat L^2/2\mu R^2 +
\hat{h}_1 + \hat{h}_2 \nonumber\\
& + \hat{V}(R,\xi)] \times \Phi_j(\xi)\,\d\xi. \label{eqWij}
\end{align}
The different equations are coupled by the off-diagonal terms $W_{ij}(R)$ with $i\ne j$.

The coupled equations may be expressed in matrix notation,
\begin{equation}
\frac{\d^2\boldsymbol{\psi}}{\d R^2}= \left[{\bf W}(R)-{\cal E}{\bf I}\right]\boldsymbol{\psi}(R). \label{eqcp}
\end{equation}
If there are $N$ basis functions included in the expansion (\ref{eqexp}), $\boldsymbol{\psi}(R)$ is a column vector of order $N$ with elements $\psi_{j}(R)$, ${\bf I}$ is the $N\times N$ unit matrix, and ${\bf W}(R)$ is an $N\times N$ interaction matrix with elements $W_{ij}(R)$.

In general there are $N$ linearly independent solution vectors $\boldsymbol{\psi}(R)$ that satisfy the Schr\"o\-ding\-er equation subject to the boundary condition that $\boldsymbol{\psi}(R)\rightarrow0$ in the classically forbidden region at short range. These $N$ column vectors form a wave function matrix $\boldsymbol{\Psi}(R)$.

\section{Uncertainties in fitted parameters} \label{app:uncer}

Our objective is to fit a set of $M$ parameters $p_j$, collectively represented by the vector $\boldsymbol{p}$, to a set of $N$ observables $y_i^\textrm{obs}$.
We minimize the weighted sum of squares of residuals,
\begin{equation}
\chi^2 = \sum_i \left( \frac{y_i^\textrm{obs}-y_i^\textrm{calc}(\boldsymbol{p})}{u_i} \right)^2,
\end{equation}
where $u_i$ is an uncertainty for observable $i$. In standard least-squares methods, with $N\gg M$, the common uncertainty of the measurements is usually estimated statistically from the minimum value of $\chi^2$ achieved in the fit, generally with a denominator $N-M$. In the present work, $N=M=4$, so this is not possible. Instead we choose the values $u_i$ as the experimental uncertainties.

To estimate uncertainties in the fitted parameters, we follow the usual procedures for non-linear least-squares fitting. At the final values of the parameters, we calculate a $4\times4$ Jacobian matrix {\bf J} with elements $J_{ij} = \partial y_i^\textrm{calc}/ \partial p_j$. We scale this by the chosen uncertainties to define the matrix {\bf A} with elements $A_{ij}=J_{ij}/u_i$ and the Hessian matrix ${\bf H}={\bf A}^T{\bf A}$; the elements of the latter are half the second partial derivatives of $\chi^2$ with respect to potential parameters. We choose uncertainties in the parameters defined by a contour at $\chi^2=1$. The variance-covariance matrix is then $\boldsymbol{\Theta}={\bf H}^{-1}$. The resulting correlated uncertainties $\Theta_{jj}^{1/2}$ are $\pm 0.006$~\AA\ in $R_\textrm{SR,0}$, $\pm 0.0016$~\AA\ in $R_\textrm{SR,1}$, $\pm 4 \times 10^3$~cm$^{-1}$~\AA$^6$ in $C_6$ and $\pm 5 \times 10^5$~cm$^{-1}$~\AA$^8$ in $C_8$.
The correlation matrix has elements $C_{ij}=\Theta_{ij}/(\Theta_{ii}\Theta_{jj})^\frac{1}{2}$; all elements have magnitude below 0.6, except that between $R_\textrm{SR,0}$ and $R_\textrm{SR,1}$, which is $-0.995$.

It should be noted that these uncertainties do not take account of model-dependence due to fixing the parameters of the mid-range potential. These are hard to estimate in a systematic way because ref.\ \citenum{Docenko:2006} did not discuss uncertainties in the parameters or the correlations between them.

In correlated fits, it often not sufficient to specify parameters to within their uncertainties. The sensitivity of calculated properties to the parameters depends on the Hessian matrix ${\bf H}$, rather than its inverse $\boldsymbol{\Theta}$ \cite{LeRoy:1998}. In order to reproduce the observables to within their uncertainties, each parameter must be specified to a precision of at least $H_{jj}^{-1/2}$, which may be much smaller than $\Theta_{jj}^{1/2}$.
The parameters in Table \ref{tab:parameter} are given to a precision based on these values.


\bibliographystyle{long_bib}
\bibliography{../all}

\end{document}